\documentclass{elsart}
\pdfoutput=1

\usepackage{graphicx}
\usepackage{amssymb}

\begin{document}

\begin{frontmatter}



\title{The G0 Experiment: Apparatus for Parity-Violating
Electron Scattering Measurements at Forward and Backward Angles}

\author[18]{D.~Androi\'c},
\author[3]{D.~S.~Armstrong},
\author[5]{J.~Arvieux\thanksref{deceased}},
\author[17]{R.~Asaturyan\thanksref{deceased}},
\author[3]{T.~D.~Averett},
\author[3]{S.~L.~Bailey\thanksref{bailey}},
\author[6]{G.~Batigne\thanksref{batigne}},
\author[11]{D. H. Beck\corauthref{cor1}},
\ead{dhbeck@illinois.edu}
\author[14]{E.~J.~Beise},
\author[9]{J.~Benesch},
% ...

%\vspace{0.1in.}
\address[1]{Kellogg Radiation Laboratory, California Institute of Technology, 1201 California Blvd, Pasadena CA 91125 USA}
\address[2]{Department of Physics, Carnegie Mellon University, Pittsburgh, PA 15213 USA}
\address[3]{Department of Physics, College of William and Mary, Williamsburg, VA 23187 USA}
\address[4]{Department of Physics, Hampton University, Hampton, VA 23668 USA}
\address[5]{Institut de Physique Nucleaire d'Orsay, F-91406 ORSAY-Cedex FRANCE}\nobreak
\address[19]{Jo\^zef Stefan Institute, 1000 Ljubljana, SLOVENIA}
\address[6]{
%Laboratoire de Physique Subatomique et de Cosmologie, 53 avenue des %Martyrs, 38026 Grenoble Cedex FRANCE
LPSC, Universit\'e Joseph Fourier Grenoble 1, CNRS/IN2P3, Institut Polytechnique de Grenoble, Grenoble, FRANCE}
\address[7]{Department of Physics, Louisiana Tech University, P.O. Box 3169 T.S., Ruston, LA 71272 USA}
\address[8]{Department of Physics, New Mexico State University, Las Cruces, NM 88003 USA}
\address[21]{Department of Physics and Astronomy, Ohio University, Athens, OH 45701 USA}
\address[9]{Thomas Jefferson National Accelerator Facility, 12000 Jefferson Avenue, Newport News, VA 23606 USA}
\address[10]{TRIUMF, 4004 Westbrook Mall, Vancouver, BC V6T 2A3 CANADA}
\address[11]{Loomis Laboratory of Physics, University of Illinois, 1110 West Green Street, Urbana IL 61801 USA}
\address[12]{Department of Physics and Astronomy, University of Kentucky, Lexington, KY 40506 USA}
\address[13]{Department of Physics, University of Manitoba, Winnipeg, Manitoba R3T 2N2 CANADA}
\address[14]{Department of Physics, University of Maryland, College Park, MD 20472 USA}
\address[15]{ Department of Physics, University of Northern British Columbia, 3333 University Way, Prince George,
BC V2N 4Z9 CANADA}
\address[16]{Department of Physics, Virginia Tech, Blacksburg, VA 24061 USA}
\address[20]{Department of Physics, University of Winnipeg, Winnipeg, MB R3B 2E9 CANADA}
\address[18]{Department of Physics, University of Zagreb, Zagreb HR-41001 Croatia}
\address[17]{Yerevan Physics Institute, Alikhanian Brothers 2, Yerevan 375036 ARMENIA}

\corauth[cor1]{Corresponding author.}

\thanks[deceased]{Deceased}
\thanks[bailey]{Current address: Center for Health Policy/Center for Primary Care and Outcomes Research, Stanford University, Stanford, CA}\thanks[batigne]{Current address: Laboratoire Subatech, Nantes, FRANCE}\thanks[benmokhtar]{Current address: Department of Physics, Computer Science and Engineering, Christopher Newport University, Newport News, VA}
\thanks[biselli]{Current address: Department of Physics, Fairfield University, Fairfield, CT}
\thanks[chao]{Current address: TRIUMF, Vancouver, BC, CANADA}
\thanks[clark]{Current address: Department of Physics and Astronomy, University of Pittsburgh, Pittsburgh, PA}
\thanks[coppens]{Current address: Department of Physics, Carleton University, Ottawa, ON, CANADA}
\thanks[covrig]{Current address: Jefferson Lab, Newport News, VA}
\thanks[cowley]{Current address: Department of Physics, Stellenbosch University, Stellenbosch, SOUTH AFRICA}
\thanks[dale]{Current address: Department of Physics, Idaho State University, Pocatello, ID}
\thanks[forest]{Current address: Department of Physics, Idaho State University, Pocatello, ID}
\thanks[grimm]{Current address: Department of Physics, Louisiana Tech University, Ruston, LA}
\thanks[guillard]{Current address: LPC, Clermont-Ferrand, FRANCE}
\thanks[guillon]{Current address: LPC-Caen, Caen, FRANCE}
\thanks[guler]{Current address: Laboratoire Leprince-Ringuet, Ecole Polytechnique, Palaiseau, FRANCE}
\thanks[gustafsson]{Current address: National Board of Patents and Registration, Helsinki, FINLAND}
\thanks[hasty]{Current address: Illinois Emergency Management Agency, Springfield, IL}
\thanks[hawthorneallen]{Current address: Department of Physica Sciences, Concord University, Athens, WV}
\thanks[horn]{Current address: Department of Physics, The Catholic University of America, Washington, DC}
\thanks[ito]{Current address: Los Alamos National Laboratory, Los Alamos, NM}
\thanks[kammel]{Current address: University of Washington, Seattle, WA}
%\thanks[king]{Current address: Department of Physics and Astronomy, Ohio %University, Athens, OH}
\thanks[kolarkar]{Current address: Department of Physics, Boston University, Boston, MA}
\thanks[kuhn]{Current address: Greens Farms Academy, Fairfield, CT}
\thanks[liu]{Current address: Department of Physics, Shanghai Jiao Tong University, Shanghai, CHINA}
\thanks[maclachlan]{Current address: Department of Physics, The George Washington University, Washington, DC}
\thanks[mammei]{Current address: Department of Physics, University of Massachusetts, Amherst, MA}
\thanks[mckee]{Current address: Department of Physics, Kansas State University, Manhattan, KS}
\thanks[mckeown]{Current address: Jefferson Lab, Newport News, VA}
\thanks[micherdzinska]{Current address: Department of Physics, The George Washington University, Washington, DC}
\thanks[moffit]{Current address: Jefferson Lab, Newport News, VA}
\thanks[muether]{Current address: Fermilab, Batavia, IL}
\thanks[nakahara]{Current address: Department of Physics, University of Maryland, College Park, MD}
\thanks[neveling]{Current address: iThemba Labs, Stellenbosch University, Stellenbosch, SOUTH AFRICA}
\thanks[nilsson]{Current address: Visma Spcs AB, V\"axj\"o, SWEDEN}
\thanks[phillips]{Current address: Department of Physics, University of New Hampshire, Durham, NH}
\thanks[pillot]{Current address: Laboratoire Subatech, Nantes, FRANCE}\thanks[quemener]{Current address: LPC-Caen, Caen, FRANCE}
\thanks[rauf]{Current address: COMSATS, Islamabad, Pakistan}
%\thanks[roche]{Current address: Department of Physics and Astronomy, Ohio %University, Athens, OH}
\thanks[rutledge]{Current address: SAIC, San Diego, CA}
\thanks[schaub]{Current address: Department of Physics and Astronomy, Valparaiso University, Valparaiso, IN}
\thanks[secrest]{Current address: Department of Chemistry and Physics, Armstrong Atlantic State University, Savannah, GA}
\thanks[spayde]{Current address: Department of Physics, Hendrix College, Conway, AR}
\thanks[suleiman]{Current address: Jefferson Lab, Newport News, VA}
\thanks[tieulent]{Current address: IPNL, Lyon, FRANCE}
\thanks[versteegen]{Current address: Department of Physics, Universit\'e de Bordeaux, Bordeaux, FRANCE}
\thanks[zeps]{Current address: Department of Physics and Astronomy, Bluegrass Community and Technical College, Lexington, KY}

\begin{abstract}
In the G0 experiment, performed at Jefferson Lab, the parity-violating elastic scattering of
electrons from protons and quasi-elastic
scattering from deuterons is measured in order to determine the
neutral weak currents of the nucleon. Asymmetries 
%of order
as small as 1 part per
million in the scattering of a polarized electron beam are
determined using a dedicated apparatus.  It consists of
specialized beam-monitoring and control systems, a cryogenic
hydrogen (or deuterium) target, and a superconducting, toroidal magnetic
spectrometer equipped with plastic scintillation and aerogel
\v Cerenkov detectors, as well as fast readout electronics for the
measurement of individual events.  The overall design and
performance of this experimental system is discussed.
\end{abstract}


\end{frontmatter}
\clearpage

%DSA \linenumbers
%\modulolinenumbers[5]

% main text

\section{Introduction}
\label{sec:introduction}

In the G0 experiment, the parity-violating elastic scattering of
electrons from protons and quasi-elastic
scattering from deuterons is measured in order to determine the
neutral weak currents of the nucleon~\cite{BEC01}.  By combining
information from the ordinary electromagnetic form factors of the
nucleon with the analogous quantities that define their neutral
weak currents, the contributions of the three lightest flavors of
quark to these currents can be extracted~\cite{MCK89,BEC89}.  The
parity-violating contribution to elastic scattering from the
nucleon, in the momentum transfer region of interest, is of order
$10^{-5}$ times the electromagnetic contribution, 
%hence a very precise experiment is required. 
and the goal is to measure it with an uncertainty of a few percent of its value.  
The G0 experiment 
in Hall C at the Thomas Jefferson National Accelerator Facility 
(Jefferson Lab) in Newport News, VA  %%%DSA 
utilizes a specialized apparatus to perform these measurements.

The experimental quantity of interest is the parity-violating
asymmetry -- the relative difference in the scattering probability
of right- and left-handed (longitudinally polarized) electrons
\begin{equation}
A_{PV} = \frac{\sigma_+ - \sigma_-}{\sigma_+ + \sigma_-}.
\end{equation}
This asymmetry is sensitive to the interference of the
electromagnetic and neutral weak currents, hence contains products
of electromagnetic and neutral weak form factors.  The experiment
is carried out in two parts, corresponding to two different
kinematic conditions: electrons scattered in the forward and
backward hemispheres, respectively.  These two measurements,
similar to the measurements in the standard Rosenbluth separation
method in ordinary electron scattering, allow us to isolate two
linear combinations of form factor products in the asymmetry
\begin{equation}
A_{PV} = \left[- G_F Q^2 \over 4 \sqrt{2} \pi \alpha \right]
{{
\varepsilon G^{\gamma}_{{E}} G^{Z}_
{{E}} + \tau G^{\gamma}_{{M}}
G^{Z}_{{M}} - (1-4 \sin^2 \theta_W )
\varepsilon^{\prime} G^{\gamma}_{{M}} G^{e}_{A}}   \over
{\varepsilon (G^\gamma_
{{E}})^2 + \tau (G_{{M}})^2}},
\label{eqn:asym}
\end{equation}
where $\alpha$ and $G_F$ are the electromagnetic and weak coupling
constants, $Q^2>0$ is the four-momentum transfer, $\theta$ is the
laboratory electron scattering angle, $G_E^{\gamma (Z)}$ and
$G_M^{\gamma (Z)}$ are the electromagnetic (analogous neutral
weak) charge and magnetic proton form factors, $G_A^e$ is the
proton axial form factor as seen in parity-violating electron
scattering (related to $G_A$ measured in elastic neutrino
scattering and the nucleon anapole moment~\cite{BEC01}), and
\begin{eqnarray}
\tau & = & {{Q^2} \over {4 M_N^2}}, \nonumber \\
\varepsilon & = & {1 \over 1 + 2(1 + \tau)\tan^2{\theta \over 2}}, \ \hbox{and} \nonumber \\
\varepsilon^{\prime} & = & \sqrt{\tau (1+\tau) (1- \varepsilon^2)}
\end{eqnarray}
are standard kinematic quantities depending on the nucleon mass
$M_N$. The two products so determined are nominally $G_E^\gamma
G_E^Z$ and $G_M^\gamma G_M^Z$.  The third term in the numerator of
Eqn.~\ref{eqn:asym}, however, also contributes to the measured
asymmetry, especially at backward scattering angles where
$\varepsilon^{\prime}$ is large. The axial form factor $G^e_A$ is
of considerable interest in and of itself. A third measurement is
therefore needed in order to isolate this term. This can be done
by measuring quasi-elastic scattering from the deuteron at
backward electron scattering angles and utilizing the charge
symmetry~\cite{MIL98} of the proton-neutron system
\begin{equation}
G^{p,n}= \frac{1}{2}\left ( G^{T=0}\pm G^{T=1} \right )
\end{equation}
where $T$ is the nucleon isospin.  These three measurements,
together with the known electromagnetic form factors of the
nucleon, $G^\gamma$, allow us to obtain separately contributions
of the three lightest quark flavors to the structure of the
nucleon~\cite{BEC01}.  The first results of both parts of the experiment are published elsewhere~\cite{PRL05,Androic}.

The primary design consideration in the experiment is the small
asymmetry to be measured.  In addition, capabilities for forward
and backward angle measurements with both hydrogen and deuterium
targets are required.  This leads to a design with a
large-acceptance magnetic spectrometer that detects recoil protons for
the forward angle measurement and electrons for the backward angle
measurements, and includes capabilities for background rejection.
Measurements at the forward angle require more novel techniques, 
hence the paper has more emphasis on that aspect of the apparatus
and technique.
The following discussion opens with a more detailed assessment of
the design requirements, and is followed by sections on the
individual subsystems: beam, target, magnet, detectors, readout
and data-acquisition electronics.  It should be noted that in such
experiments control and monitoring of the beam is critical, therefore
we also include a section on the relevant aspects of the
accelerator system.  More detailed descriptions of the
target~\cite{COV04} and forward angle electronics~\cite{Marchand} can be found
elsewhere.

\section{Summary of Design Requirements}
\subsection{General considerations}                                           
The asymmetries to be measured over the desired range of momentum
transfers, $0.1 < Q^2 < 1$ GeV$^2$, range
from about 1 part-per-million (ppm) for the lowest $Q^2$ forward
measurement to about 40~ppm for the highest $Q^2$ at backward angles
(all asymmetries, as defined above, are negative).  In order to
measure these asymmetries with a relative precision of a few
percent, statistical and systematic uncertainties must be very
small (generally $< 10^{-7}$).

\begin{figure}[tbp]
\begin{center}
{\includegraphics[width=5.5in]{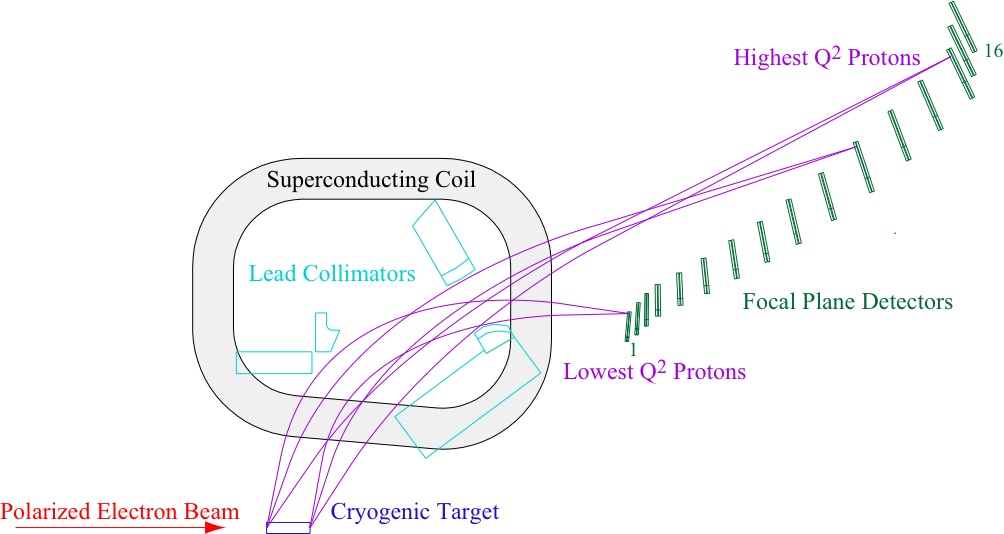}}
\end{center}
\caption{\label{fig:trajectory} Schematic of some typical trajectories
for elastic protons at different $Q^2$ in the forward measurement. 
A superconducting coil outline and one octant of collimators, line-of-sight shielding, and focal plane detectors are pictured. Protons
of the same $Q^2$ originating from any location in the target are focused
to a common detector.  Note that the highest $Q^2$ protons appear in FPD 14 -- see text.  A photo of the FPDs appears in Fig.~\ref{fig:FPD}}
\end{figure}

The statistical uncertainty ($\Delta A = 1/\sqrt{N_{tot}}$, where
$N_{tot}$ is the total number of detected events) is attained
 using a combination of high luminosity and large solid angle
acceptance.  For the G0 experiment, beam currents of $20-60 \mu$A are
used with a 20 cm liquid hydrogen (or deuterium) target for a
luminosity of about $1-3\times 10^{38}$ cm$^{-2}$s$^{-1}$.  The
spectrometer has a solid angle acceptance of roughly 0.9~sr and corresponding momentum
acceptances for elastic scattering.  These acceptances comprise the full range of momentum transfer in the
forward measurement, and that corresponding to a single momentum transfer in the
backward measurement as described below.  Further, the acceptance is
flat along the length of the target and the optics are such that
particles from different points along the target are focused to
the same location on the focal surface (see Fig.~\ref{fig:trajectory}).

The small systematic uncertainties achieved derive mainly
from the photocathode source of polarized electrons in the accelerator gun;
the remarkably precise control of the intensity, pointing and
polarization of the laser used to drive the source allows the
helicity-correlated changes in electron beam parameters to be very
small.  For example, the electron beam intensity
is controlled with an active feedback system which adjusts the
light incident on the photocathode, such that the
helicity-correlated electron beam intensity variations at the
target average to a few ppm in a typical one hour long run.  In addition, the
spectrometer is azimuthally symmetric (8 sectors -- ``octants''),
so any small helicity-correlated beam motion effects cancel to
lowest order when summing the rate over all the
detectors.

\begin{figure}[tbp]
\begin{center}
{\includegraphics[width=5.5in]{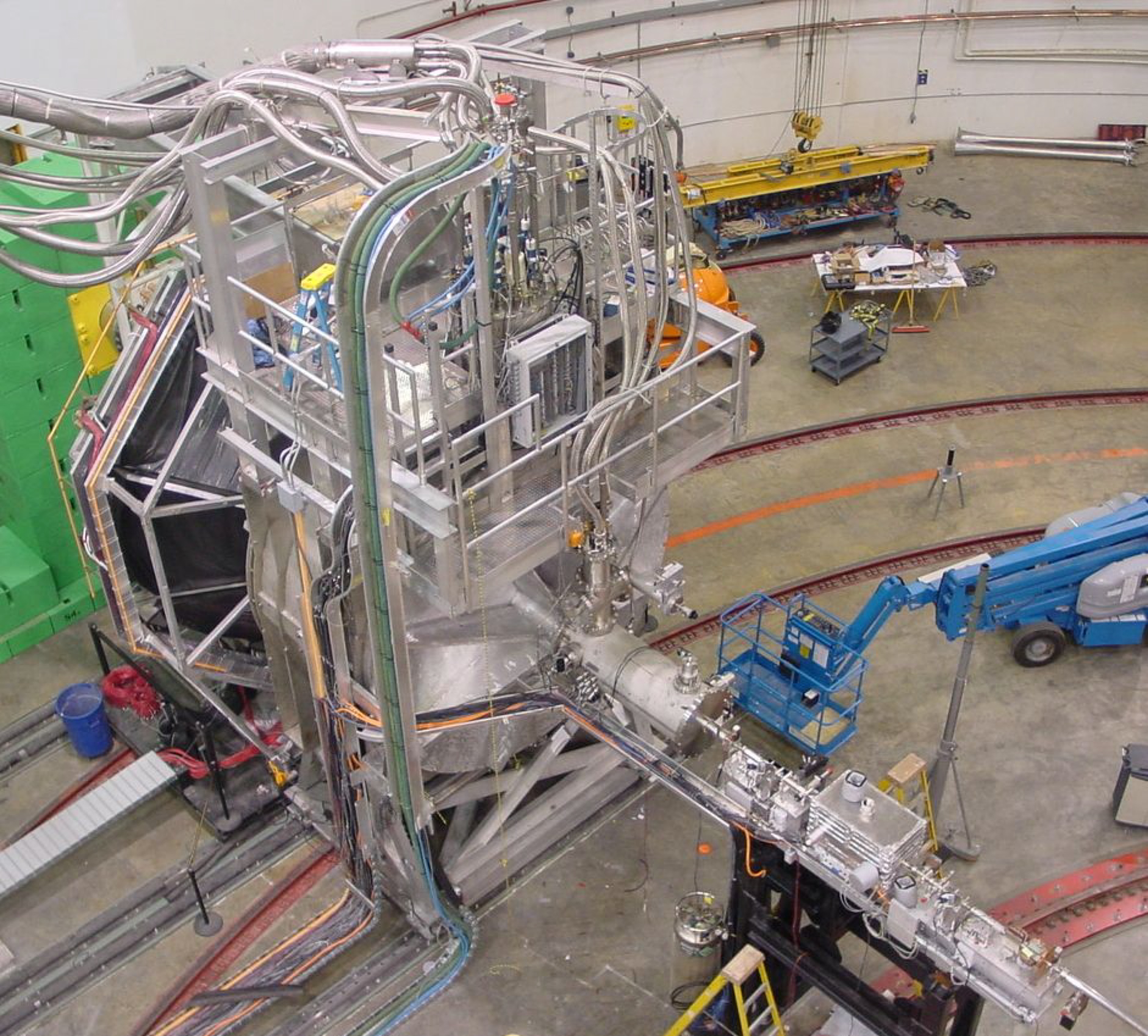}}
\end{center}
\caption{\label{fig:forwardPhoto} A view of the experimental setup in the forward configuration.  The electron beam travels from lower right to upper left.  The superconducting magnet is seen just to the left of the center of the frame; the target is centered inside the magnet.  Recoil protons from elastic scattering are bent by the magnet to the octants of focal plane detectors (FPDs) whose black covers are visible at left (Octant 7, in the horizontal plane, is the most visible in this image).}
\end{figure}

\begin{figure}[tbp]
\begin{center}
{\includegraphics[width=5.5in]{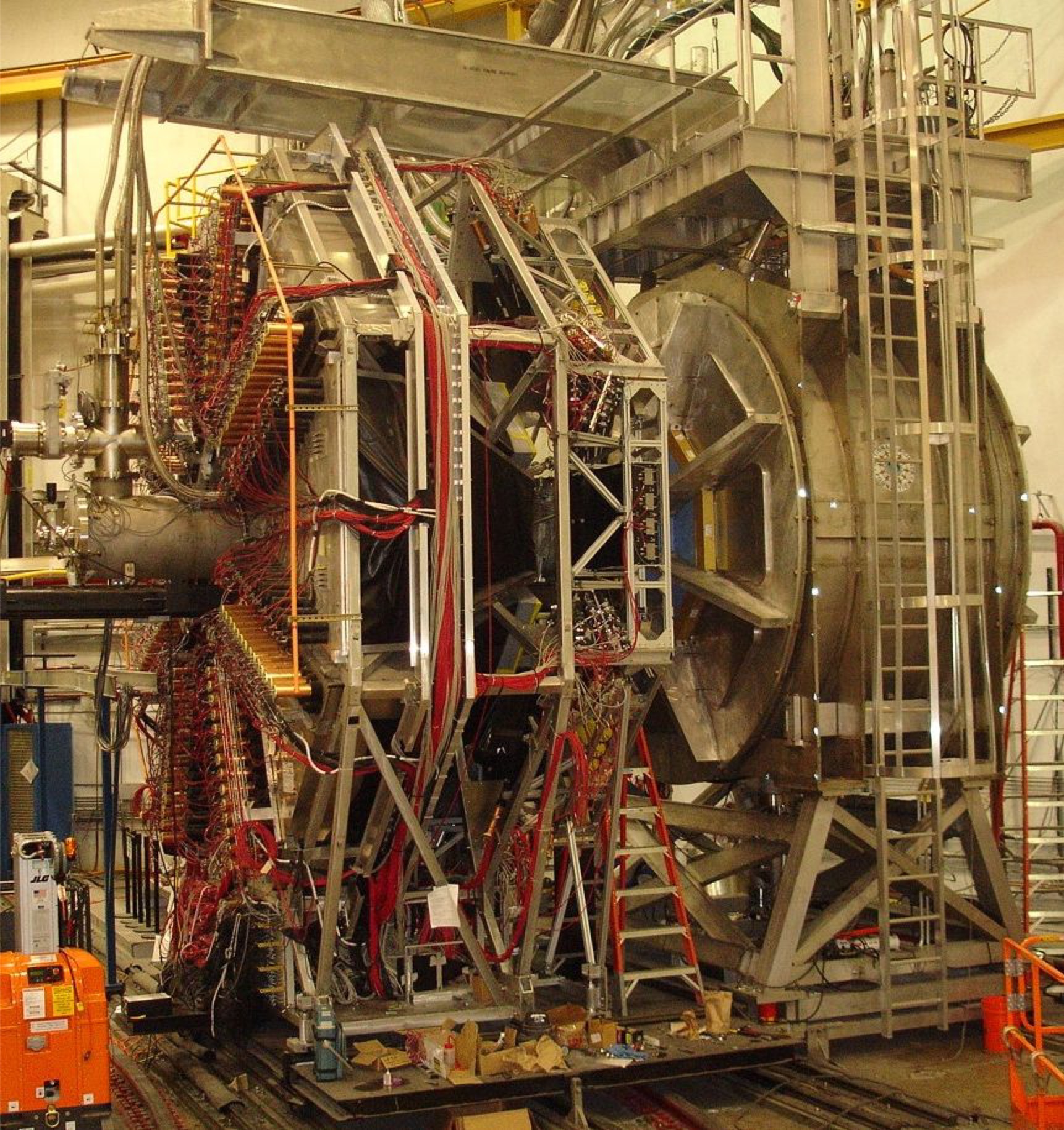}}
\end{center}
\caption{\label{fig:backwardPhoto} A view of the experimental setup in the backward configuration.  The beam enters from the left.  The superconducting magnet is on the right of the frame; the target is centered within it.  Electrons scattered backward are bent by the magnet into the detector system (shown retracted from its normal position by about 1 m) in the center-left region of the image.  The photomultiplier tubes for Octant 3 are closest to the camera at left.  Detector supports added for the backward measurement, holding the Cherenkov and cryostat exit detectors (CEDs), are visible just to the right of center.}
\end{figure}

As indicated above, a complete description of the neutral weak
currents of the nucleon requires measurements at both forward and
backward angles.  A spectrometer with modest $\int |\vec B| 
dl$ can be used if one detects the
recoil protons for the forward measurement and the scattered
electrons for the backward measurement.  Further, detection of
``forward'' protons at about $70^\circ$ and ``backward'' electrons
at the complementary angle of $110^\circ$ (in separate
measurements) allows for reasonable kinematic leverage to separate
the $G_E^\gamma G_E^Z$ and $G_M^\gamma G_M^Z$ terms.  Thus by
reversing the spectrometer with respect to the beam direction,
both measurements can be performed with basically the same
instrument.  Photographs of the experimental setup in the forward and backward configurations are shown in Figs.~\ref{fig:forwardPhoto} and \ref{fig:backwardPhoto}. A summary of the parameters for the 
experiment is presented in Table~\ref{tab:kinematics}.  With an
acceptance that ranges between roughly $52^\circ$ and $77^\circ$
in the forward direction 
and a beam energy of 3~GeV, the full
range of momentum transfers can be measured in a single setting.
The corresponding proton momenta vary between about 0.35 and
1.13 ~GeV/c, and the corresponding scattering angles of the 
(undetected) electrons range from $6^{\circ}$ to $21^{\circ}$. 
For the same range of momentum transfers, the electron
momenta at the backward scattering angle of about $110^\circ$ are
also within this range.
However, because $Q^2$ varies slowly with angle in the backward
direction, separate measurements are made to sample the desired 
range of momentum transfer; beam energies of  
359 and 684~MeV gave $Q^2 = 0.22$ and $0.63$ GeV$^2$, respectively.  
Background processes are rejected using different methods
for the forward and backward measurements, as described below.

\begin{table}[tbp]
\begin{center}
\begin{tabular}{lll}
Quantity & Forward & Backward \\
& protons & electrons \\ \hline
Beam energy & 3.0 GeV & 0.36, 0.69 GeV\\
Target length & 20 cm & 20 cm\\
Beam current & 40 $\mu$A & 20-60 $\mu$A\\
$p$ & 0.35 - 1.13 GeV/c & 0.24, 0.35 GeV/c\\
$\theta_{elastic}$ & 52.0 - 76.5$^\circ$ & 100 - 118$^\circ$\\
$Q^2$ & 0.12 - 1.0 GeV$^2$ & 0.22, 0.63 GeV$^2$\\
$\Delta \phi$ & $0.44\cdot 2\pi$ & $0.44\cdot 2\pi$ \\
$\Delta \Omega_{elastic}$ & 1.07 sr & 0.82 sr \\
$\int |\vec B| dl$ & 1.6 T $\cdot$ m & 1.6 T $\cdot$ m \\
\end{tabular}
\end{center}
\caption{Nominal parameters for the forward and backward measurements which
illustrate the general capabilities of the G0 apparatus.  The scattering angle acceptances are different
for forward and backward measurements because of the differing correlation between scattering angle and momentum
for the two cases.}
\label{tab:kinematics}
\end{table}

\subsection{Forward angle measurement}\label{ss:forwarddesign}                                           
The choices of detector and electronics for the experiment are
dictated by the relatively high rates in the forward measurement.
The superconducting magnet system (SMS) 
is segmented in octants with corresponding
detectors; the detector widths within each octant are chosen to
give reasonable resolution in forward momentum transfer as well as to
limit the rate to $\sim $2~MHz (with about 1/2 coming from elastic
protons, 1/2 coming from background inelastic protons and pions).
Pairs of plastic scintillators with phototubes at each end are
chosen for the detector elements.  The basic elastic proton event
is defined by a coincidence of the left-right mean-timed signals from two
such scintillator elements.

In order to separate elastic protons from background in the
forward measurement (without explicitly measuring trajectories),
time-of-flight (ToF) measurements are used to determine particle
momenta.  The combination of spectrometer optics and kinematics
leads to a situation where inelastic protons and pions in a given
detector have higher momenta than the elastic protons and thus
appear at earlier times (see also Fig.~\ref{fig:tof}).  
In order to be able to measure the
roughly 20~ns ToF of the elastic protons, the beam
for this experiment has 32~ns between electron
bunches (in contrast to the usual 2~ns spacing for each
experimental hall) using a 31.1875~MHz pulsed laser to operate the
electron source.  With this arrangement, pions appear at the
detectors about 7~ns after the beam passes through the target;
inelastic protons arrive after the pions and before elastic
protons. ToF spectra are incremented for every beam
pulse with custom time-encoding electronics (see Section~\ref{sec:electronics}).  Care must be taken with ToF start signals, which are
generated by passage of the beam bunch through the target, to ensure,
for example, they are not helicity-correlated (see also
Section~\ref{sec:beam}).

The optics of the magnet are such that the range of $Q^2$ for the forward measurement is
dispersed along the focal surface.
As the recoiling elastic proton angle decreases (and its momentum
increases), the trajectory intersects the focal surface further
from the beam axis, as illustrated in Fig.~\ref{fig:trajectory}. 
This allows a separation of $Q^2$ simply by
segmenting each octant into a number of separate detector
elements.  The combination of the rate limit and reasonable $Q^2$
resolution leads us to 15 detector elements per octant (``focal
plane detectors'' or FPDs).  At the top of the focal surface the
proton trajectories ``turn around'' with higher $Q^2$ protons
starting to again move closer to the beam axis.  In the top two
detectors, additional $Q^2$ resolution is obtained by separating
the elastic protons by ToF.  A 16$^{th}$ detector is used at the
very top of the focal surface to help monitor backgrounds.

\subsection{Backward angle measurement}\label{ss:backwarddesign}                                         

The backward angle measurements are qualitatively different for
two reasons: electrons are detected and the range of $Q^2$
available in a single setting is small.  Because the electrons are ultra-relativistic,
ToF cannot be used to separate inelastic and elastic electrons.
Therefore, the more traditional approach of trajectory measurement is used
for the backward angle measurements.  A set of small scintillators
is placed near the exit window of the magnet (``cryostat exit
detectors'' or CEDs) in each octant.  Certain combinations of CED
and FPD detectors correspond to the correct combination of
momentum and angle for the elastic electrons; other combinations
correspond to inelastic electrons (see also Fig.~\ref{fig:matrix}).  For the backward angle
measurements, the mean-timed signals from the FPDs\footnote{In order to reduce low energy backgrounds, the mean-timed signal
for a given FPD detector element (consisting of a front/back scintillator pair) 
is formed by using one tube (at opposite ends) of each of the front and back FPD.} are
combined with the corresponding signals from the CEDs (using a
programmable logic array) to sort the events.  As
with the FPDs, the phototubes for the CEDs are located in a low
magnetic field region using long light-guides.

Negative pions pose a different kind of problem for the backward
angle measurements, as they cannot be
distinguished using the CED-FPD detector pairs described above.  An
aerogel \v Cerenkov counter with a threshold pion momentum of about
570~MeV/c, above the range of $\pi^-$ momenta accepted, is used to
eliminate them from the electron trigger.  This counter
is placed between the CEDs and the FPDs with the phototubes again
located in a relatively low field region.  The \v Cerenkov is most
important in the measurement with the deuterium target where
quasi-free $\pi^-$ production from neutrons yields $\pi/e$ ratios
of order 100 at the highest momentum transfers.  (For the
hydrogen target, the $\pi^-$ are essentially all produced in the
aluminum target windows.)

\subsection{Counting rate effects}                                         

Because individual detector events are counted rather than integrated
in this experiment (integration is more common in parity-violation
measurements because of the high rates~\cite{SAMPLE,HAPPEX}), rate corrections 
are important in both the forward (deadtime) and backward angle 
(deadtime and accidental coincidences) measurements.  
They are corrected directly from beam current dependence of the trigger rates and from singles
measurements; helicity-correlated beam current variation effects 
are corrected together with other helicity-correlated parameters such as beam position
(see also Sections~\ref{sec:beam} and \ref{sec:electronics}).

\section{Polarized Beam}
\label{sec:beam}

Like all parity-violating electron scattering experiments, the
G0 experiment requires a polarized electron beam of high
intensity and the ability to control and accurately
measure its properties.  These requirements are driven
by both the statistical and systematic error goals of the
experiment.  The high figure of merit ($P^2 I$) needed to achieve
a small statistical error is obtained with a high current ($I
\sim 20-60\ \mu$A) and high polarization ($P > 70\%$) electron beam.
The small asymmetries demand that careful attention be
paid to the helicity-correlated properties of the beam.  These
properties must be measured and controlled well enough so that any
correction made to the measured asymmetries is
relatively small and well-understood.  Achieving these goals requires
specialized subsystems and a degree of coordination between the
experimental and accelerator operations that are not typical of
more conventional electron scattering experiments.  In addition to
the 
specialized polarized beam needs typical of
parity-violating electron scattering experiments, the 
forward-angle portion of the 
G0
experiment also requires a time structure 
different from
the usual time structure of the CEBAF accelerator 
at Jefferson Lab.  
%These issues
%are covered in some detail in the following sections.

\subsection{Polarized source}

The polarized source
of the CEBAF accelerator is based on the
photoemission from GaAs photocathodes of polarized electrons which
are accelerated to 100 kV in an electron ``gun''.  Detailed descriptions
of the polarized source and past performance can be found in
Ref.~\cite{SIN07}.  For the G0 experiment, strained GaAs
photocathodes~\cite{SPI04} are used.  These crystals typically
provide electron beam polarizations 
of  
$\sim 70 - 85\%$ at the
desired currents.  The backward angle measurement utilized so-called superlattice
strained cathodes, yielding the polarizations in the 85\% range~\cite{MAR04}.
Laser light
from up to three different drive lasers (for the three CEBAF
experimental halls) is incident on the
photocathode through a vacuum window.  The drive lasers emit
radio-frequency pulsed light with typically $\sim$100~ps pulse
%Chk Sinclair paper for width
widths, synchronous with subharmonics of the CEBAF accelerating
frequency of 1497 MHz. The electrode structure of the gun then
focuses and accelerates the electrons to a kinetic energy of 100
keV.  The standard repetition rate for beam delivered to a single
experimental hall from the CEBAF accelerator is 499 MHz~\cite{LEE01}; this is the mode used for backward angle
G0 running.  

To make
use of the ToF technique for forward angle running,
the G0 experiment requires a lower repetition rate (31.1875
MHz), giving a 32~ns spacing between beam pulses. %%DSA
To maintain 
high average current at the 
%%%DSA  low 
lower repetition rate, a Time-Bandwidth Products~\cite{TBW05} Ti-Sapphire laser
providing $\>$ 300 mW at 840 nm is used.  It is important
that the random noise in both the intensity and direction of the 
laser light be small
at the helicity-reversal frequency of the experiment.
In practice,
the r.m.s. spread in beam parameters in the
experimental hall, as determined from 
measurements made at the 30 Hz helicity reversal frequency,
is always $<$~$0.1 \% $ for intensity and $< 10$~$\mu$m for the
centroid of the beam position.

The other significant injector issue for forward angle running
is beam transport. 
The current requirements (40 $\mu$A at 31 MHz) imply a peak
charge of 1.3~pC per micropulse, in comparison to the more typical
value of 0.2 pC (100 $\mu$A at 499 MHz) for usual high-current
operation.  Using standard settings, pulses with this high peak charge suffer
significant emittance growth due to space charge effects in the 10
meter long 100 keV region,  leading to significant transmission losses and
poor quality beam. 
An
acceptable solution was achieved with several modifications to the
injector region hardware and tuning procedures, including adding
new magnets, modifying typical laser parameters, and stabilizing
RF systems.  A tune was developed that satisfied
the requirements of G0 and those of a simultaneously running 
experiment in Hall A~\cite{hypernuclear} that
required a very small fractional energy spread ($2.5 \times
10^{-5}$).  Complete details of these developments can be found 
elsewhere~\cite{KAZ04}.

\subsection{Beam monitoring and control}
\label{sec:beam_mon}

For the G0 experiment, it is important to have continuous
monitoring of essential beam properties so
corrections for helicity-correlated beam properties can be
made. When the beam helicity is reversed, ideally, no 
other %%DSA
property of the beam would change.
In reality, many properties of the
beam, such as position, angle, and current are observed to
change.  This causes the false asymmetry

\begin{equation}
\label{eq:Afalse} A_{false} = \sum_{i=1}^{N}
\frac{1}{2 Y} \frac{\partial Y}{\partial P_{i}}
{\Delta P_{i}},
\end{equation}

where $Y$ is the number of detected scattering events normalized to beam current, 
%%DSA
the $P_{i}$ are the beam
properties (position, angle, current, and energy), and the
$\Delta P_{i} = P_{i}^{+} - P_{i}^{-}$ are the helicity-correlated differences
of those beam properties.  Generally, these false asymmetries are
kept small by careful beam setup or by active feedback as described below.  In this
section, we discuss beam monitoring and the beam controls necessary 
for measurement of both the yield slopes
($\partial Y / \partial P_{i}$) and the correlated differences ($\Delta P_{i}$).

The beam current is measured with two microwave cavity monitors
installed in the Hall C beamline approximately 40 m upstream
of the G0 target.  The monitors are cylindrical, stainless
steel cavities resonant in the TM$_{010}$ mode at 1497 MHz and with a
$Q \sim 500$.  Beam electrons passing through the cavity excite a
resonance; the energy is extracted from an antenna installed in
the cavity.  This is a convenient resonant mode for measuring the
beam current because the spatial dependence of the electric field
amplitude is nearly constant in the center of the cavity, thus
making the response relatively insensitive to beam position. The high frequency signal
is mixed down to a DC level, which
is measured using a voltage to frequency converter and a 
%%%DSA   scalar
scaler
that is read out at the same rate as the rest of the data in the
experiment.  The intrinsic noise level in the monitors at the 30 Hz helicity-reversal
rate is measured (by comparing two nearby monitors) to be $~40$~ppm (compared to typical
beam current fluctuations of  500 - 1000 ppm).

The beam position is measured at many points along the Hall C
beamline with ``stripline'' beam position monitors~\cite{BAR91}.
These monitors consist of a set of four thin wires placed
symmetrically around the beam; each wire has a length of one
quarter wavelength at 1497 MHz. Beam power coupled into the four
antennae at 1497 MHz is downconverted to a lower frequency,
filtered, and converted to a DC voltage.  The monitors are
outfitted with 
switched electrode
electronics~\cite{POW97}, supplemented with special sample
and hold modules to provide a signal for each antenna. The
resulting voltage signals are sent through voltage to frequency
converters and recorded with scalers read out at the same rate as
the rest of the data in the experiment. Linear
combinations of antenna signals are used to extract the beam
positions in software.  The electron beam position and angle
projected to the 
target are determined with a pair of
stripline %%%DSA 
monitors separated by 2.5 m with a midpoint 4.8 m
upstream of the target.  The electron beam energy is determined
from the in-plane beam position measured by a similar monitor located at the center
of the Hall C arc, where the beam dispersion is 40~mm/\%.  
The intrinsic noise level in these monitors at the 30 Hz helicity-reversal
rate is measured (again by comparing two nearby monitors) to be 3 $\mu$m (compared to typical
beam position fluctuations of  $\sim$ 10 $\mu$m).

Beam ``halo'' is a 
generic term for electrons associated with the incident
beam that are radially far ($> 10\ \sigma$) from the main beam.
Significant beam halo can interact with an 11~mm diameter flange that
is part of the G0 target cell.  To minimize this interaction
our 
beam specification requires that $< 1 \times 10^{-6}$
of the electrons in the beam are located outside a 3~mm radius from
the center of the beam.  The amount of beam halo is continuously monitored
using a 2~mm thick aluminum target with a 6~mm diameter hole in it,
located 8 m upstream of the G0 target.  Scattered
particles from beam interactions with this ``halo target''
are detected at large ($\sim 15^{\circ}$)
and small ($\sim 3^{\circ}$) angles.  The detectors consist of
5 cm photomultiplier tubes attached to small pieces of lucite or
scintillator.  The 
system
is calibrated by directing a 5~nA
beam into the 2~mm thick aluminum frame of the halo
target.  With the normal beam tune it is possible to achieve the
halo specification on a routine basis (corresponding to negligible rates in the
G0 spectrometer detectors).  The halo monitor is useful as
a diagnostic of significant changes in the beam tune that require corrective
action.  

Measurements of the yield slopes, $\partial Y / \partial P_{i}$, are made by 
deliberately varying the beam parameters, or by using the natural variations thereof.
Beam position and angle at the target are varied
periodically 
using a set of six air-core
steering coils positioned 
downstream 
of the Hall C dispersive arc.  The
positions of the coils are chosen 
to insure the space of $x$ and $y$ positions and angles at the
target is adequately spanned (typical ranges of $\pm 0.5 \ \hbox{mm}$ and
 $\pm 0.5 \ \hbox{mr}$, respectively).  
The beam
energy and current
are periodically 
modulated by controlling the power input to an accelerating
cavity in the CEBAF accelerator's South Linac and the beam current using polarized source
intensity control (described below), respectively.

\subsection{Helicity control electronics}\label{helicity_control}
\begin{figure}[tbp]
\begin{center}
{\includegraphics[width=5.5in]{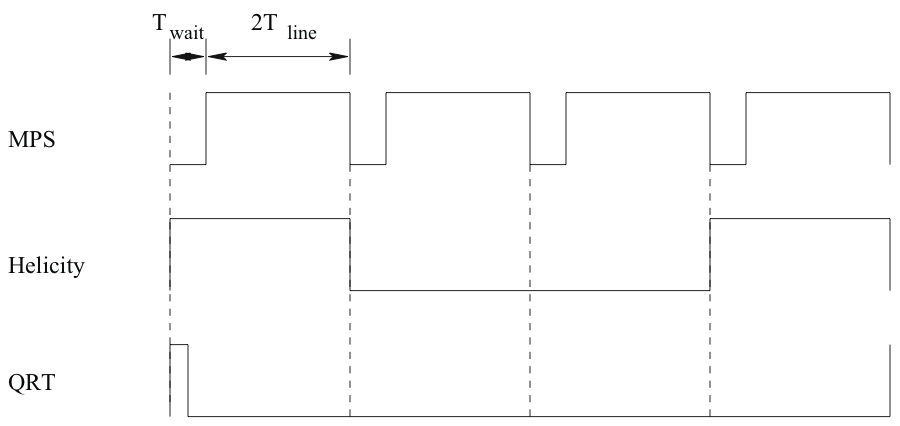}}
\end{center}
\caption{Timing scheme of the signals from the helicity control 
electronics. T$_{\rm line}$ = 1/60 s is the power line period.}
\label{fig:g0_hel_new}
\end{figure}

The helicity of the electron beam is set by the polarization of laser
light incident on the strained GaAs photocathode.  The polarization is controlled by a
Pockels cell driven by high voltage power supplies set to correspond
to $\pm \lambda/4$ phase retardation, or approximately
$\pm 2.5$ kV at our nominal laser wavelength.  
A high voltage switch, driven by a digital ``helicity signal'', determines which supply drives
the Pockels cell.
The helicity control electronics, built in a single VME module around a field programmable gate array 
(FPGA), generates various timing signals, including that for helicity, as shown in  
Fig.~\ref{fig:g0_hel_new}.

The macropulse trigger (MPS) signal is the primary data-taking
integration gate for the experiment and is set to twice the period of
the 60 Hz line power.  
All signals in the experiment are integrated over
this period, allowing for the dominant 60 Hz noise (and that of the highest subharmonic)
in the electron beam properties and electronic signals
to be averaged out.  The MPS ``off'' state, typically 500
$\mu$s ($T_{wait}$), is set to allow the Pockels
cell to stabilize after a helicity change.  During this period
the experiment takes no data, and all the scalers are read out.

The beam helicity is changed relatively rapidly
to reduce the effect of slow drifts.  
To provide exact cancelation of linear
drifts in an asymmetry measurement, 
we %%%DSA
generate the helicity signals in a quartet (``QRT'') sequence.  The initial helicity
state of a sequence is chosen randomly by a pseudorandom number
generator programmed into the FPGA.  This initial helicity state determines
the rest of the QRT sequence, $+ - - +$ or $- + + -$, and generates a corresponding control
signal for the acquisition electronics.

All signals from the helicity control electronics are delivered
to the acquisition electronics via fiber optic cable to insure 
complete ground isolation, and eliminate a possible source of false
asymmetries.
Other cross-talk between the helicity control electronics and the acquisition
electronics is suppressed by 
delaying the helicity signal reported
to the acquisition electronics by eight MPS signals relative
to the actual helicity signal sent to the Pockels cell
high voltage switch.  The true helicity is reconstructed in
software from the delayed information and cross-checked with knowledge of the 
pseudorandom pattern.

\subsection{Beam position and intensity control} 

As indicated in Eqn.~\ref{eq:Afalse}, it is very important
to minimize the helicity-correlation in
beam 
intensity, position, angle, and energy.
This is accomplished by careful setup of both the polarized injector laser beam and the accelerator optics to minimize helicity-correlation, as well as by active feedback to the injector.

\begin{figure}[tbp]
\begin{center}
\includegraphics[width=5.5in]{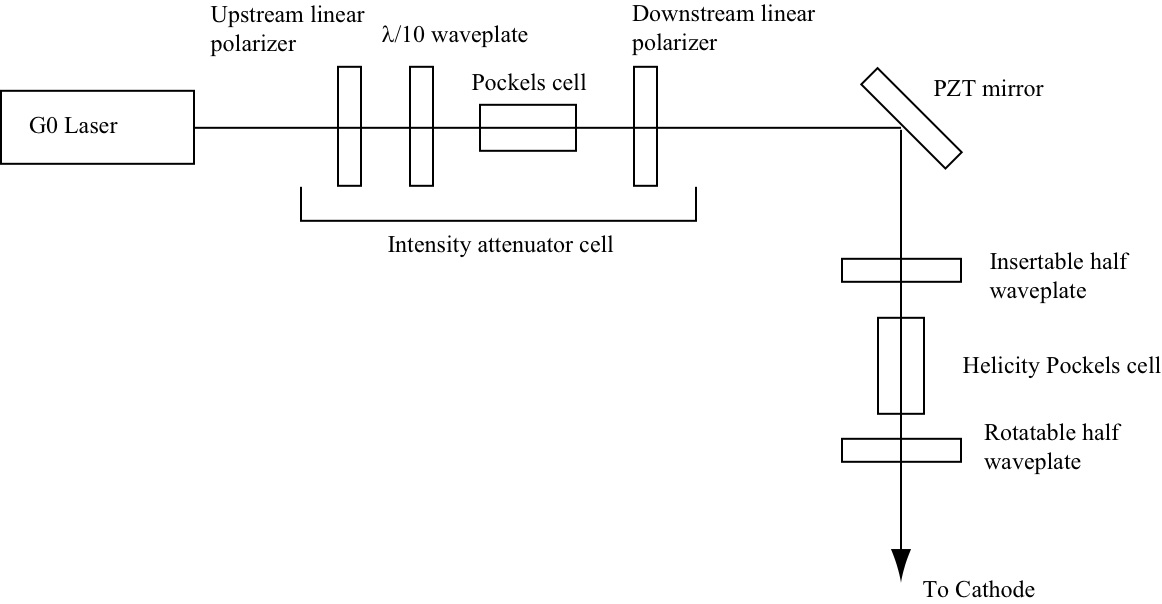}
\end{center}
\caption{A schematic illustration of the important devices associated
with the minimization of helicity-correlated effects in the beam
current and position.}
\label{fig:feedback_devices}
\end{figure}

One of the most important sources of helicity-correlated intensity and position differences
in the electron beam is the Pockels cell that sets the laser
helicity.  The small residual
linear polarization component produced by the cell (different for the two nominal circular polarization states) is transported to the crystal with an efficiency dependent on its orientation.  The result is
helicity-correlated variations in the laser intensity; a similar
effect arising from different steering of the laser beam for the two polarization
states is partly responsible for the helicity-correlated position differences.
In addition to the interaction with the optical elements, the overall ``analyzing power'' of
the strained GaAs photocathode plays an important role.  The photocathode quantum efficiency 
depends on the orientation of the residual linear polarization~\cite{MAI96}, and can, in addition,
have substantial overall gradients.  Depending
on the state of the system, this can result in rather substantial
intensity asymmetries ($\sim$ 10,000 ppm) and position differences
($\sim$ 10,000 nm).

\begin{figure}[tbp]
\begin{center}
\includegraphics[width=5.5in]{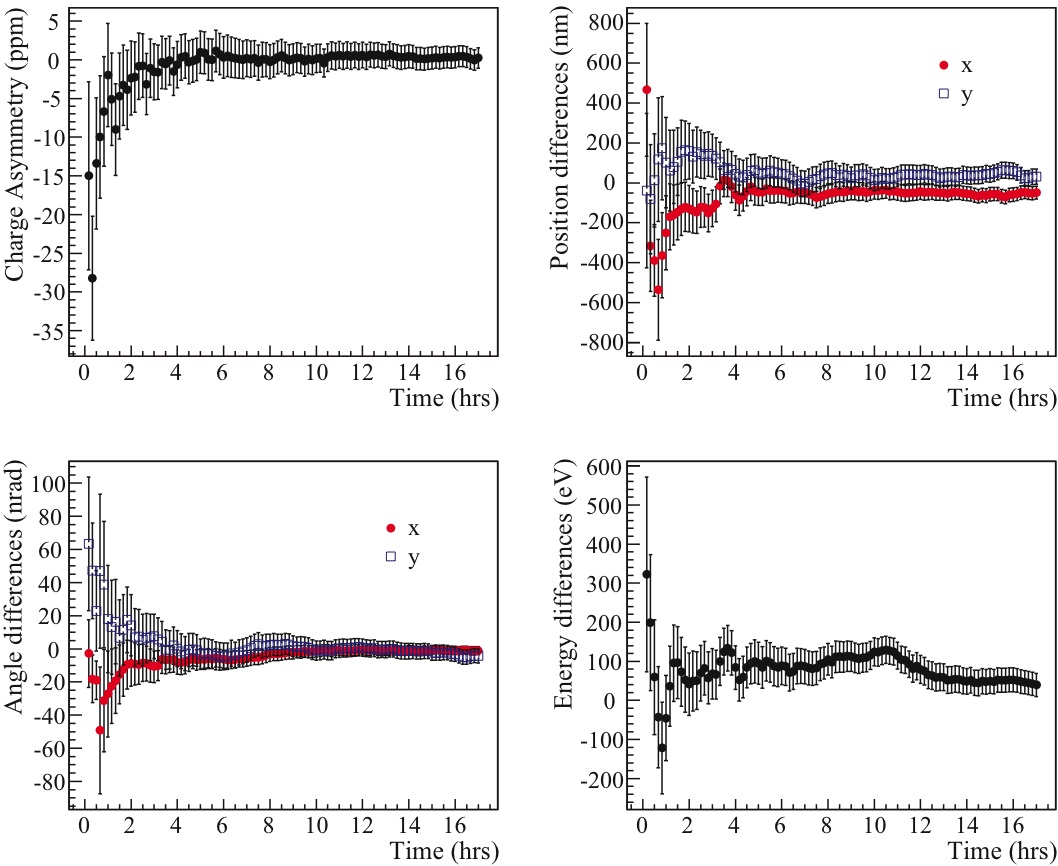}
\end{center}
\caption{Feedback results (forward angle) with both the
charge and position feedback circuits operating. The cumulative averages
of the charge asymmetry
(upper left), horizontal and vertical position differences (upper
right), horizontal and vertical angle differences (lower left) and
energy differences (lower right) are plotted versus time from the
start of a feedback cycle. As these are running averages, successive data points
are highly correlated.}  
\label{fig:feedback_results}
\end{figure}

The setup of the accelerator system to achieve small helicity-correlations in beam
properties involves three main elements.
First, a rotatable $\lambda$/2 waveplate is inserted into the laser line just downstream
of the helicity Pockels cell~\cite{HAP04} to rotate the residual linear
component to minimize the intensity asymmetry and position differences.
Second, care is taken in the accelerator tuning to realize the
natural ``adiabatic damping'' in the acceleration process. For an
ideally tuned accelerator, the transverse position variations of the beam are
reduced 
$\propto 1/\sqrt{p}$ where $p$ is the beam momentum.  Due to
imperfections in the electron beam transport, the full suppression
is never achieved, but suppression factors of 10-25 are
observed between position differences in the experimental
hall versus those in the 
injector region. 
After these two passive measures, the
intensity asymmetries observed at the 
target are
typically  $<$ 100 ppm, while the position differences are
typically  $<$ 300 nm.  In both the forward and backward angle measurements,
active feedback is used to improve the intensity asymmetries.  For the backward angle measurements,
it is found that more careful alignment of the optical elements (especially that of the
helicity Pockels cell) could reduce the position differences to $\sim 20$ nm; for the forward
angle measurements, position differences are reduced by active feedback.
Intensity asymmetries are reduced with
feedback to the IA (intensity attenuator)
cell (Fig.~\ref{fig:feedback_devices}).
It consists of two linear polarizers oriented parallel to each other
with a Pockels cell in between.  The Pockels cell is operated at low, helicity-dependent
voltages (0 - 50 volts) to change the transmission of the attenuator.\footnote{The $\lambda$/10 waveplate
is necessary to insure a non-zero slope in the intensity versus
voltage relationship when the Pockels cell is operating at low voltage.}
During regular running, the intensity asymmetry is measured for three minutes
(during which time it is typically measured with a precision of 10 - 20
ppm) and then corrected in an automated cycle (see Fig.~\ref{fig:feedback_results}).

Helicity-correlated beam position differences are corrected during forward angle
running by feeding back to a piezoelectric-controlled steering mirror which moves 
the laser beam on the photocathode in a helicity-correlated manner
(Fig.~\ref{fig:feedback_devices}).
In this case, the beam position differences are
measured and corrected on a 30 minute cycle (using measurements with
a precision 
of 100-200 nm, see Fig.~\ref{fig:feedback_results}).  In practice, it is found that these intensity and beam position ($x$ and $y$)
controls are not completely
orthogonal; a full 3x3 matrix system is required for successful 
feedback~\cite{NAK04,NAK06}.  

\begin{figure}[tbp]
\begin{center}
\includegraphics[width=5.5in]{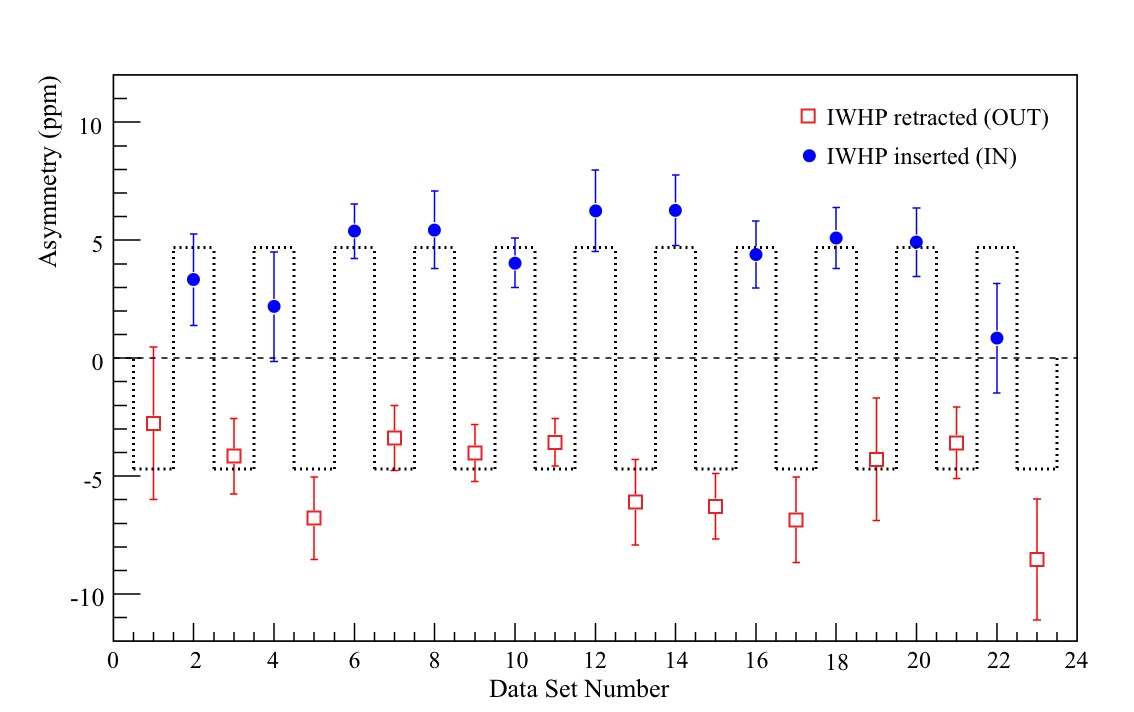}
\end{center}
\caption{Asymmetries of forward angle elastic protons in FPD 9. The squares (circles) correspond to the data with the insertable half-wave plate retracted (inserted). The measurements correspond to an average asymmetry of $-4.72\pm0.32$ ppm ($\chi^2/d.o.f. = 18.0/22$), shown in the figure as the dotted line (with the appropriate changes of sign).}
\label{fig:oscillation}
\end{figure}

\begin{figure}[tbp]
\begin{center}
\includegraphics[width=5.5in]{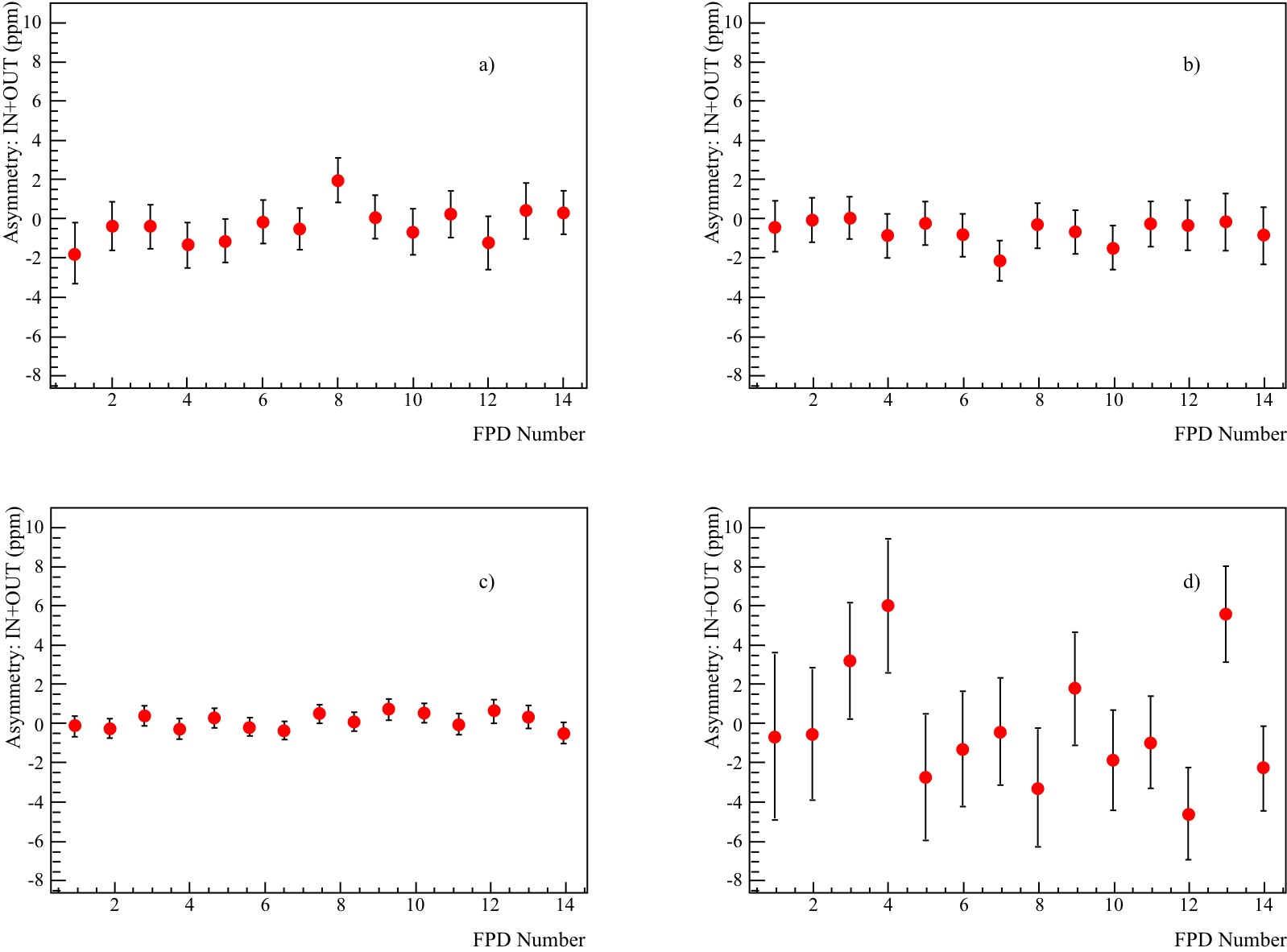}
\end{center}
\caption{Sum
of the forward angle asymmetries with the IHWP inserted (IN) and retracted (OUT) for each FPD; a) and b) for ToF windows immediately preceding the elastic window, c) elastic window and d) window following the elastic window.  The average asymmetry (ppm) and $\chi^2/\nu$ are: $-0.28\pm0.31$, 0.58; $-0.64\pm0.31$, 0.29; $0.02\pm0.12$, 0.65; and $-0.45\pm0.74$, 1.34, for a), b), c) and d), respectively.  The widths of the time windows for a) and b) are half that of the elastic window; window d) has the same width as that of the elastic.}
\label{fig:IN+OUT}
\end{figure}

Lastly, Fig.~\ref{fig:feedback_devices} also shows the insertable half-wave plate (IHWP) used to reverse the sense of the beam helicity relative to all electronic signals in the polarized source (the half-wave plate changes right-circular polarization to left-, and vice versa).
%  The position of this half-wave plate (``IN'' or ``OUT'') is recorded only by hand in the experimental logbook.  
The measured asymmetries reverse as expected with the IHWP ``OUT'' to ``IN'' (helicity reversed) changes as illustrated in Figs.~\ref{fig:oscillation} and \ref{fig:IN+OUT}, showing no indication of contributions from electronic crosstalk.  The asymmetry for FPD 9 (forward angle) is shown in the first figure for the 23 separate data-taking periods with different IHWP positions demonstrating good statistical consistency.  The overall consistency and reversal dependence for four different ToF regions (forward angle) is shown in the IN + OUT summed asymmetries in Fig.~\ref{fig:IN+OUT}.

\subsection{Polarimetry}%%%added by DSA

The electron beam polarization is measured periodically using the Hall
C M{\o}ller polarimeterA~\cite{Moeller}.  The polarimeter measures the asymmetry in
the cross section for electron-electron scattering, for which the
analyzing power is accurately known. The target electrons are provided
by a pure iron foil saturated in a 3 T field.  The scattered and
recoiling M{\o}ller electrons produced at 90$^{\circ}$ in the center of
mass are detected in coincidence using a symmetric apparatus. The beam
polarization is typically measured with $<1$\% statistical error in 5 min.
For the forward angle data-taking, the
average beam polarization over the entire data set is measured
to be $73.7 \pm 1.0$\% where the uncertainty is dominated by the systematic uncertainty
in extrapolation from the low beam current of the polarization measurement to the operating
current of the main data-taking~\cite{Phillips}.  The polarization for the backward angle experiment is $85.8 \pm 2.1 (1.4)\%$
at 359 MeV (684 MeV), again dominated by systematic uncertainties.  For the higher energy, in addition to the beam
current extrapolation uncertainty, there is an additional contribution of comparable
size due to beam tuning effects.  For the low energy running, the M{\o}ller polarimeter could not be operated; instead a 5 MeV Mott polarimeter in the 
injector region, calibrated using the M{\o}ller measurement at 684 MeV (and incorporating knowledge of the beam spin transport~\cite{Grames}), is used
to monitor the polarization~\cite{Price}.  The uncertainty in this calibration is the origin of the additional factor in the overall low-energy
polarization uncertainty.

\subsection{Raster}

The intrinsic diameter of the electron beam ($\sim 200 \; \mu$m) is such
that at the currents used in these measurements, $20 - 60$~$\mu$A, the power 
density on the cryogenic target would be 
$\sim 1$~kW/mm$^2$. 
To reduce this power density, and therefore the magnitude of beam-induced density fluctuations
in the target (see Section~\ref{targetperformance}),
the beam is rastered using two magnets located approximately
20 m upstream of the target. The fast raster system~\cite{raster}
generates a square, $\sim 2 \times 2$~mm$^2$ pattern by sweeping 
the fields in $x$,$y$ with triangular
waveforms of 24.96 kHz and 25.08 kHz frequency, respectively,
yielding a pattern with 95\% uniformity in the beam density.                                                        

\section{Target System}                                                  
\label{sec:target}

The target system consists of the main cryogenic target, filled with either liquid hydrogen or liquid deuterium, as well as auxiliary solid targets for diagnostic purposes.  In order to determine the backgrounds from the aluminum target cell, the main cryogenic target could also be operated with gas at a temperature several degrees above that used for liquid operation.  The auxiliary solid targets are C and Al; a ``halo'' target (actually just a 5.6 mm diameter hole -- distinct from the upstream halo monitor target -- see Section~\ref{sec:beam_mon}) is also included.  Finally, as discussed below, an additional thin aluminum target could be placed in the beam  just downstream of the cryogenic cell to better characterize the downstream window of the cell (see Fig.~\ref{fig:flyswatter}).

\subsection{Cryogenic target}

The G$0$ cryogenic target is a closed-loop recirculating system for liquid
hydrogen and deuterium,
with a 20~cm long target cell that resides inside the
vacuum enclosure of the superconducting magnet. 
%%%DSA added below.
Here we provide an overview of the target and its performance;
a more detailed description is available elsewhere~\cite{COV04,COV05}. 

Design considerations for the target include the beam power requirements
($\sim 340$~W at 60~$\mu$A beam current), space constraints due
to the 
60~cm
diameter opening in the magnet, non-magnetic construction for the parts in
the magnetic field, and the 
need for reliable operation for many months of operation 
with no servicing.  
As a result, the cryogenic loop is oriented horizontally.

The target fluid is circulated with a velocity $\sim$m/s through the cryogenic loop in order to dissipate the power
deposited by the electron beam as it passes through the target cell. The
fluid is pumped by a Barber Nichols~\cite{BarberNichols} custom DC brushless,
``sensorless'' motor, driven by a sensorless controller. A tachometer,
consisting of a Cu coil and a small permanent magnet, measures the pump
rotation frequency. The
variable voltage induced by the changing magnetic flux through
the Cu coil as the motor shaft rotates is monitored by a
%n Agilent 34401A 
digital multimeter enabled in frequency mode. The nominal operating
frequency of the motor is 30~Hz; the maximum available torque
in liquid hydrogen is found to be 0.16 N-m at 42.7~Hz.

On the opposite side of the cryogenic loop is a counterflow heat
exchanger with finned Cu tubing on the target fluid side.  The
coolant side uses helium gas at 15K and 12 atm from the Jefferson Lab End
Station Refrigerator (ESR).  The effective area for heat exchange
is 9501~cm$^2$ on the target side and 1110~cm$^2$ on the coolant
side, with a measured heat transfer coefficient of 214~W/K under
normal operating conditions. Tests of the maximum performance of
the heat exchanger found it capable of removing up to 1000~W of
heat from the liquid hydrogen at its nominal operating point (19~K
and 1.7~atm) when using 80\% of the available coolant from the ESR.

After passing through the heat exchanger, the target fluid enters
a target manifold to which the thin-walled target cell is
soldered. 
It holds both the primary hydrogen cell
and a secondary cell filled with helium that serves as the
entrance window to the hydrogen system.  The primary cell is a
thin-walled cylinder machined from one piece of Al-7075, of
length 23~cm and inner diameter 5~cm. An exit window spot, centered
on the beam axis, is machined to a 
thickness 0.076~mm and diameter 8~mm; the remainder of the shell is
approximately 0.178~mm thick. Target cells are pressure
tested to 590 kPa.  The target fluid is directed longitudinally
down the center of the target cell by a thin-walled inner cone.  
Small holes in the side of the cone produce turbulence
and allow for mixing in the longitudinal flow. The 16~cm long helium cell ($\sim 0.07$ g/cm$^2$)
upstream of the primary cell is attached to the manifold
with a flange and indium seal. The downstream window of the
helium cell, soldered to the main body, is 0.228~mm thick and
serves as the entrance window to the hydrogen.  To reduce variations in target
length for differing beam positions, the radius of curvature of this window
is matched to that of the primary exit window; the helium cell is
maintained at the same pressure and temperature as the hydrogen
cell. At its upstream end, the helium cell has a 0.178~mm thick Al window. The
length of liquid hydrogen seen by the detector system is
20~cm or 1.44 g/cm$^2$ at the operating conditions of 19~K and 170 kPa
; the net material in the
three windows seen by the detector is 0.130~g/cm$^2$.  In addition, for studies of backgrounds arising from these 
entrance and exit windows, the cryotarget is
also successfully operated with hydrogen gas at 220 kPa pressure 
at two operating temperatures, 28 K and 33 K.

The cryogenic loop is supported inside the SMS vacuum system
on a cantilevered platform connected to a service module upstream of
the SMS. The cantilever and service module are designed and
constructed by Thermionics~\cite{Thermionics}. The location of the target with respect
to the incident beam is controlled by four actuators (two vertical,
two horizontal) which provide pitch, yaw, and translation of the
entire cryogenic target loop vertically and horizontally. The target is
aligned to
the magnet-beam axis to $<1$~mm
horizontally and vertically.

The control system for the cryogenic target has the primary
functions of monitoring and periodically recording the various
target parameters, providing warnings of critical conditions, and, using
a proportional-integral-differential (PID) feedback system,
maintaining the target fluid at a constant temperature. It is
based on a VME processor, 
%Greenspring ADIO
standard I/O modules, and the EPICS~\cite{EPICS} slow controls system.  The temperature of the target fluid is
monitored in six locations within the cryogenic loop with
Lakeshore Cernox CX-1070-AA~\cite{Lakeshore} resistors immersed directly in the
fluid. Two additional sensors monitor the temperature of the
helium coolant at each end of the heat exchanger. Target pressure
is monitored in the gas system. Pressure excursions in
the target loop are minimized by a 9450 l ballast tank
coupled to the gas-handling system. The temperature of the target
is maintained with a resistive heater located just
downstream of the target manifold before the target fluid enters
the circulating pump. The heater is constructed of three
coils of Ni-Cr alloy ribbon, each with a resistance of
3.5~$\Omega$, wired in parallel,
% to give a net resistance of 1.15~$\Omega$, 
and driven by a 40~V, 25~A DC power supply.
During normal operations, the total heat load on the target is held
constant at approximately 400~W, coming from either the incident
beam or from the heater. The feedback system tracks the beam
current incident on the target, subtracts the deposited beam power
from the target power, and sets the current on the
heater to make up the deficit. Temperature excursions, 
even after a beam trip, are less than 0.2~K.

\subsection{Solid targets} 

In addition to the cryogenic loop, several solid targets are
mounted on the target manifold to allow measurement of background
processes.  For example, events from the aluminum target windows typically
account for a few \% of the elastic yield in the forward angle measurement, and up
to $10-15\%$ in the backward angle measurement.

Using the position actuators, the target assembly can be pitched downwards for
an out-of-beam, no-target geometry. A dummy target frame is situated
between these two positions, and consists of a 3.2 mm thick
aluminum frame bolted to the cryogenic target manifold flange. This
frame, 13.4 cm upstream of the center of the 20 cm long cryogenic cell, 
supports three different targets just above the cryogenic cell, insuring
there is no interference 
with the trajectory of the protons or electrons scattered anywhere along the 20~cm length
of the main cryogenic target. 
The
three targets are: a 5 mm thick carbon target mounted over a
10 mm diameter hole, a bare 5.6 mm hole used to characterize the beam
halo, and a third target which simply consists of a surveyed spot on
the aluminum frame itself.

\begin{figure}[tbp]
\begin{center}
\includegraphics[width=5.5in]{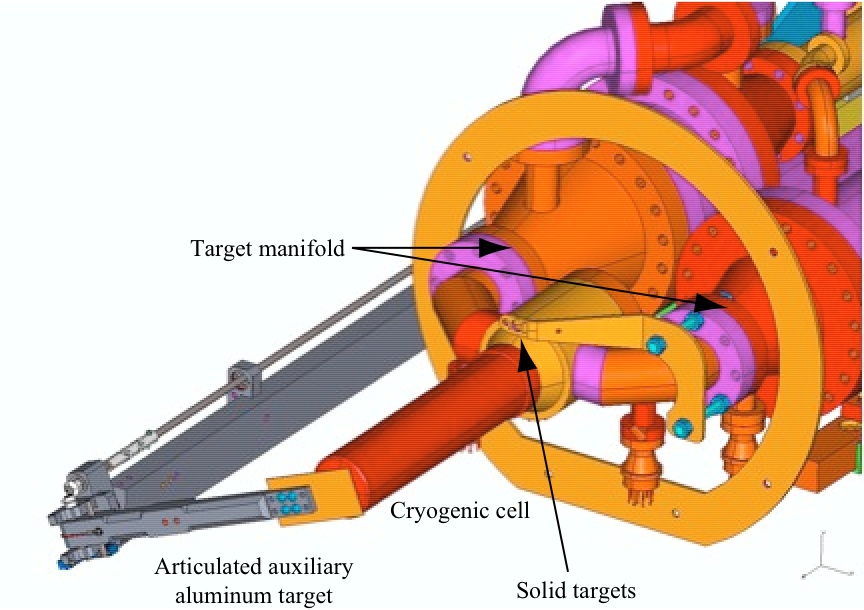}
\end{center}
\caption{Drawing of the cryogenic target, with solid target frame shown
above the cylindrical cryogenic cell, 
and with the auxiliary articulated
aluminum target downstream of the cell shown in the inserted position.  The horizontal manifold contains the pump (left) and the heat exchanger (right).  The beam enters from the upper right along the axis of the cryogenic cell. 
}
\label{fig:flyswatter}
\end{figure}

An auxiliary aluminum target is provided 1 cm downstream of the exit
window of the cryogenic cell and is used to characterize the response from
the downstream target window; see Fig.~\ref{fig:flyswatter}. 
This 0.76 mm thick Al foil can be articulated on or off the beam axis
independently of the other targets.  
It is used in conjunction with a special 0.085
mm thick tungsten foil target 48 cm upstream of the aluminum foil
target. The tungsten target is the same thickness (in radiation lengths) as
the cryogenic target, but is located well upstream of the detector acceptance. It is
used to radiate photons for Al foil target measurements, mimicking the 
background at the exit window from
photons produced within the cryogenic target.

\subsection{Luminosity monitors}  
\label{sec:lumi}

Primarily to study and monitor beam-induced variations in  
fluid density in the target, a set of luminosity monitors 
is located downstream of the target at very forward
angles~\cite{Liu}.
Short-term
target density fluctuations can contribute to the overall statistical
noise in the measurement, combining with that from the particle counting statistics. 
The experimental goal is to reduce the target density effects on the noise
to a negligible level by adjusting the operating parameters of the beam (size) and target (flow speed, etc.).

The luminosity detectors consist of
synthetic quartz {\v C}erenkov cubes coupled to low-gain
photomultiplier tubes (PMT) at an angle of approximately 2$^\circ$ with
respect to the incident beam. A total of eight detectors are placed
symmetrically about the beam line.
Due to the very high rate of electrons coming primarily 
from M{\o}ller scattering in the target, the phototube
current is integrated over the
30~Hz helicity window.  It is then converted to a voltage, passed through a
voltage-to-frequency converter, and counted with scalers. These
electronics are similar to those used for the beam position
and current monitors described in Section~\ref{sec:beam_mon}. 
The luminosity monitors
provide a significantly higher precision for small changes in
asymmetry widths, each having an intrinsic width of 200~ppm per
QRT with
a 40 $\mu$A beam current. This is compared to the asymmetry
width of each of the 15 rings of G$0$ FPD detectors of
approximately 1200~ppm in the forward measurement and a minimum of 
5000~ppm for the full elastic signal at backward angles. However, because the luminosity monitors are more
sensitive to other small systematic effects such as small changes
in beam halo or scraping of the beam upstream of the target, some care must
be taken to separate effects in determining the target density contributions
to detector signal widths.

\subsection{Target performance}\label{targetperformance}  

Prior to operation with liquid hydrogen, tests of the target system were carried out with cold helium gas in order to
establish the performance characteristics of the heat exchanger,
and with liquid neon in order to asses the safety of the gas
handling system in the event of a catastrophic failure~\cite{COV05}. 

Tests with liquid hydrogen included assessment of 
density fluctuations via measurement of the width of asymmetry
distributions as a function of beam current, target pump speed,
beam raster size~\cite{raster} and intrinsic beam spot size. For the initial, forward
angle measurement, in order to
improve the sensitivity of these tests relative to the statistical
precision of a single G$0$ detector, groups of detectors were
averaged together and integrated over ToF. In addition, the asymmetry
distributions for the luminosity monitors (Section~\ref{sec:lumi}) provided additional information. 

A detailed discussion of
these tests and analysis of the data can be found
in Ref.~\cite{COV04}. The overall conclusion 
is, for nominal running conditions with a 40~$\mu$A incident 
beam rastered over a square of side 2~mm, and with
the target circulating pump operating at approximately 30~Hz, the maximum
contribution to the asymmetry widths coming from density
fluctuations (hydrogen or deuterium) is of order 240$\pm$70~ppm. 
This results in at most a 2\%
increase in the signal width for any $Q^2$ bin in the forward
angle measurement and has a negligible impact in the backward 
angle measurements. Global density reductions
of the liquid due to beam heating were also investigated,
and are found to be less that 1.5\% at the nominal operating
conditions.

\section{Magnetic Spectrometer}                                                 
\label{sec:magnet}
The G$0$ experiment employs a superconducting magnetic spectrometer (SMS) to 
analyze the momentum
of scattered particles (protons in the forward mode, electrons in the 
backward mode), to define the angular acceptance for scattered 
particles, and to
provide both magnetic and bulk
material shielding for the detectors against backgrounds originating from the target.
By reversing the orientation of the spectrometer 
with respect to the target center, while also reversing the magnet
polarity, both the forward proton and backward electron modes are
accommodated. The additional conditions
 placed on the spectrometer magnet by the requirements of the  experiment are summarized in Table~\ref{tab:sms-req}.

\begin{table}[tbp]
\begin{center}
\begin{tabular}{l p{1.8in} p{2in}}
Requirement       & Strategy      & Implications \\ \hline
High statistics         & High luminosity       & \parbox[t]{2in}{\raggedright
                                  Extended target (20~cm length)} \\
                & High rates            & Minimal particle tracking \\
                & Large $\phi$ acceptance   & Minimal obstruction \\
\vspace{-0.1in.}
Large $Q^2$ range       & Modest $Q^2$ resolution   & $1\% \le \Delta Q^2/Q^2 \le 10\%$ \\
(Forward measurement) \\
Low backgrounds         & \parbox[t]{2in}{\raggedright
                  Reduce target background}
                                & \parbox[t]{2in}{\raggedright
                                  Line-of sight-shielding
                                  (sets minimum bend-angle, $\alpha$)} \\
Low systematics         & Axial symmetry        & Symmetric spectrometer \\
                & \parbox[t]{2in}{\raggedright
                  Minimize spin-dependent re-scattering}
                                & \parbox[t]{2in}{\raggedright
                                  Iron-free environment, low target field} \\
\end{tabular}
\end{center}
\caption{General requirements and design implications for the G0 spectrometer.\label{tab:sms-req}}
\end{table}

%Some of the critical choices made in the optical design and
%the resulting magnetic field are discussed
%in the following section. A description of the mechanical and
%electrical design is presented in Section~\ref{sec:sms-physical}.
%This is followed in Section~\ref{sec:sms-verification} by a
%description of the manufacturing tolerances implied by the design
%and the means by which they are verified.
%Section~\ref{sec:sms-controls} presents the control system with
%emphasis on its quench protection function. Finally, we conclude,
%in Section~\ref{sec:sms-performance} with comments on the
%performance of the magnet.

\subsection{Optics\label{sec:sms-optics}}

Various optical designs
of the SMS were considered, including solenoidal and various dipole/multipole configurations,
but it soon became clear that 
a toroidal spectrometer offered many advantages.
The large acceptance arising from a relatively unobstructed
geometry, and the intrinsic axial symmetry of a toroid are
particularly attractive.  In addition, there need be no iron
return yokes or pole faces, potential sources of spin-dependent
re-scattering, which could produce false asymmetries.  The magnetic field
of a toroid is negligible near the axis where the target is
located, hence beam steering and target polarization effects are
minimized. A toroidal magnet can also accommodate both the forward
proton and backward electron measurements with a simple reversal
of orientation along its axis and a corresponding reversal of the
magnet polarity.

The forward-proton measurements place the most restrictive
constraints on the spectrometer and therefore drive a number of
the design parameters. The size and scale of the spectrometer is
set by the larger proton momentum, and the minimum bend angle,
$\alpha$, of about $35^\circ$, is dictated by the requirement that the
detectors in their optimal location be shielded from direct view
of the target. The goal for the momentum transfer range of the
experiment is $0.1 \le Q^2 \le 1.0$~GeV$^2$.  
Resolution in $Q^2$ of order 10\% provides a number
of distinct measurements in this range. This corresponds, for example,
to a momentum resolution, $\Delta p/p$, of about $\pm 5\%$ at $Q^2 =
0.5$~GeV$^2$. In the backward case, it is only necessary to
resolve elastic and inelastic electrons, which requires a
resolution $\Delta p/p \ge \pm 10\%$ over the
$Q^2$ range of the experiment.

Although they provide, at best, a crude representation, first-order
TRANSPORT~\cite{transport} matrix elements, $(\hbox{detector coordinate}|\hbox{target coordinate})$, 
are useful in discussing the broad characteristics of the SMS.  Relative to
a central trajectory, the characteristic
coordinate pairs are $x,\theta$ corresponding to the dispersive direction ($\vec v \times \vec B$)
and the scattering angle, and $y,\phi$ corresponding to the azimuthal coordinates.
Because of both relatively better momentum resolution for an extended target, and
a smaller detector package (bending toward the axis), ``zero-magnification'', 
$(x|x)=0$, optics is chosen over the usual, focusing, $(x|\theta)=0$,
optics.  In the $(x|x)=0$ case, each point on the focal surface ideally corresponds to a
particular $(p,\theta)$ pair.  For the two-body final state in elastic scattering, the correlation
between $p$ and $\theta$ leads, in principle, to different values of $Q^2$ for each value of $x$ at
the focal surface.  This situation is approximately realized for the forward scattered protons where
the momenta vary rapidly with scattering angle.  For the backward scattered electrons, deviations from $(x|x)=0$
away from the octant median plane dominate the effect from the relatively much smaller variation of momentum
with scattering angle.

\begin{figure}[tbp]
\begin{center}
\includegraphics[width=5.5in]{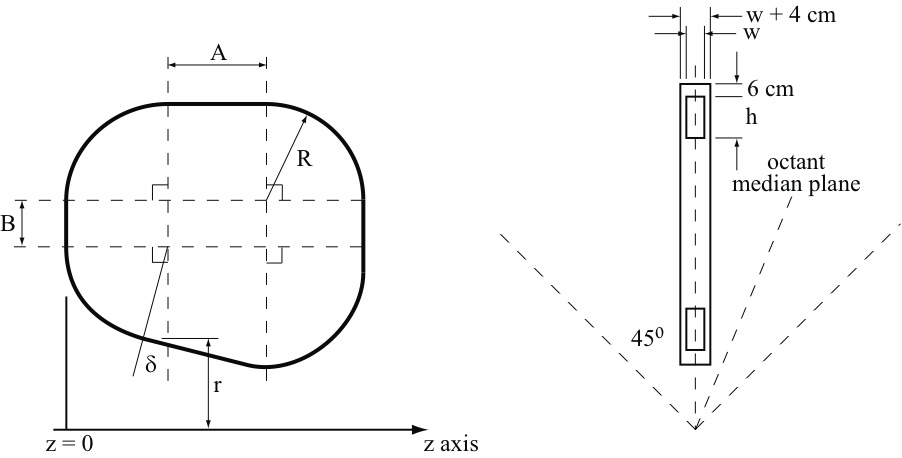}
\end{center}
\caption{
SMS coil schematic (outside dimensions).  Optimized dimensions are listed in Table~\ref{tab:sms-coil-param}.}
\label{fig:sms-coil-profile}
\end{figure}

Operating cost projections dictate that the magnet be superconducting.
Construction costs are reduced by using a common cryostat for all the coils.  The choice
of superconducting technology allows
a relatively higher number of ampere-turns near the beam axis, both decreasing
the azimuthal obstruction of the coil and allowing the rotation of the inner 'edge' 
of the magnetic field (effected by the taper of the otherwise $\sim$ rectangular coils at the inside radius as shown in Fig.~\ref{fig:sms-coil-profile}) to move the focal surface outside the cryostat.  Eight coils are
chosen as a reasonable compromise between azimuthal acceptance and optical aberrations
away from the median planes (caused by the stronger curvature of the magnetic field lines near
the coils). 

Schematic views of the coil outline and cross section are shown in Fig.~\ref{fig:sms-coil-profile}.
A conservative current density of 5~kA/cm$^2$ is chosen for the surplus Superconducting Super Collider superconducting wire used.
The parameters $h$, $w$, $r$, $R$, $A$, $B$, and
$\delta$, defined in Fig.~\ref{fig:sms-coil-profile} are varied to essentially achieve the best zero-magnification condition with the largest
solid angle acceptance, subject to a maximum cryostat diameter of 4 m (the
size of the door opening in Jefferson Lab Hall C) and the constraint that the focal surface
lie outside the cryostat.  The optimized values of these parameters are shown in Table~\ref{tab:sms-coil-param}
along with the central target position and the total coil current (5 kA x 144 turns).

\begin{table}[tbp]
\begin{center}
\begin{tabular}{cc}
Quantity & Value \\
\hline
$h$ & 18 cm \\
$w$ & 8 cm \\
$r$ & 42 cm \\
$R$ & 50 cm \\
$A$ & 65 cm \\
$B$ & 28 cm \\
$\delta$ & $­5^\circ$ \\
$z$ (target) & 44 cm \\
$I_{\rm coil}$ & 0.72 MA \\
\end{tabular}
\end{center}
\caption{\label{tab:sms-coil-param}Optimized G0 spectrometer
current distribution parameters (see text).}
\end{table}

In order to eliminate background and to ensure that the resolution
requirements are met, collimators are introduced to define
the acceptance.
In order to limit the effect of aberrations near the edges of the octants, azimuthal collimators (10 cm thick
lead along the trajectory) are used to limit the $\phi$ acceptance to about $\pm 10^\circ$.

``Primary'' lead collimators, located at a neck in the envelope of
particle trajectories (see Figs.~\ref{fig:trajectory} and \ref{fig:sms-cutaway}),
limit the acceptance in the dispersion direction.  Both the low
and high momentum limits for elastic protons are defined by the primary
collimator nearest the beamline.  This lower primary collimator is about 70 cm thick along the line-of-sight
from the target to the detectors in order to scatter neutrons effectively.  The other two pieces of the primary collimators
shown in Fig.~\ref{fig:sms-cutaway} reduce background from particles outside the desired
acceptance as well as from showering of electrons and photons generated in the target.

For the backward measurements, the momentum resolution of the
spectrometer must be sufficient to separate elastic from
threshold-inelastic scattering. The leading terms in momentum
resolution are
\begin{equation}
\label{equ:sms-res}
(\frac{\Delta P}{P})^2 = (\frac{(x|x)}{(x|\delta)})^2\cdot\Delta x^2 +
                         (\frac{(x|\theta)}{(x|\delta)})^2\cdot\Delta \theta^2
\end{equation}
Although the resolution could be improved by reducing the length of
the target, $\Delta x$, the required resolution is more readily
achieved by restricting the angular acceptance $\Delta \theta$
because of the size of the coefficients in
Eqn.~\ref{equ:sms-res}.  This restriction is effected by inserting a new set
of plastic scintillators (CEDs) near the exit of the magnet cryostat.
The combination of the FPDs and CEDs are sufficient to separate elastic and inelastically scattered
electrons, and allow us to measure both parity-violating asymmetries simultaneously~\cite{We04}.

\subsection{Physical design \label{sec:sms-physical}}
\begin{figure}[tbp]
\begin{center}
\includegraphics[width=4in]{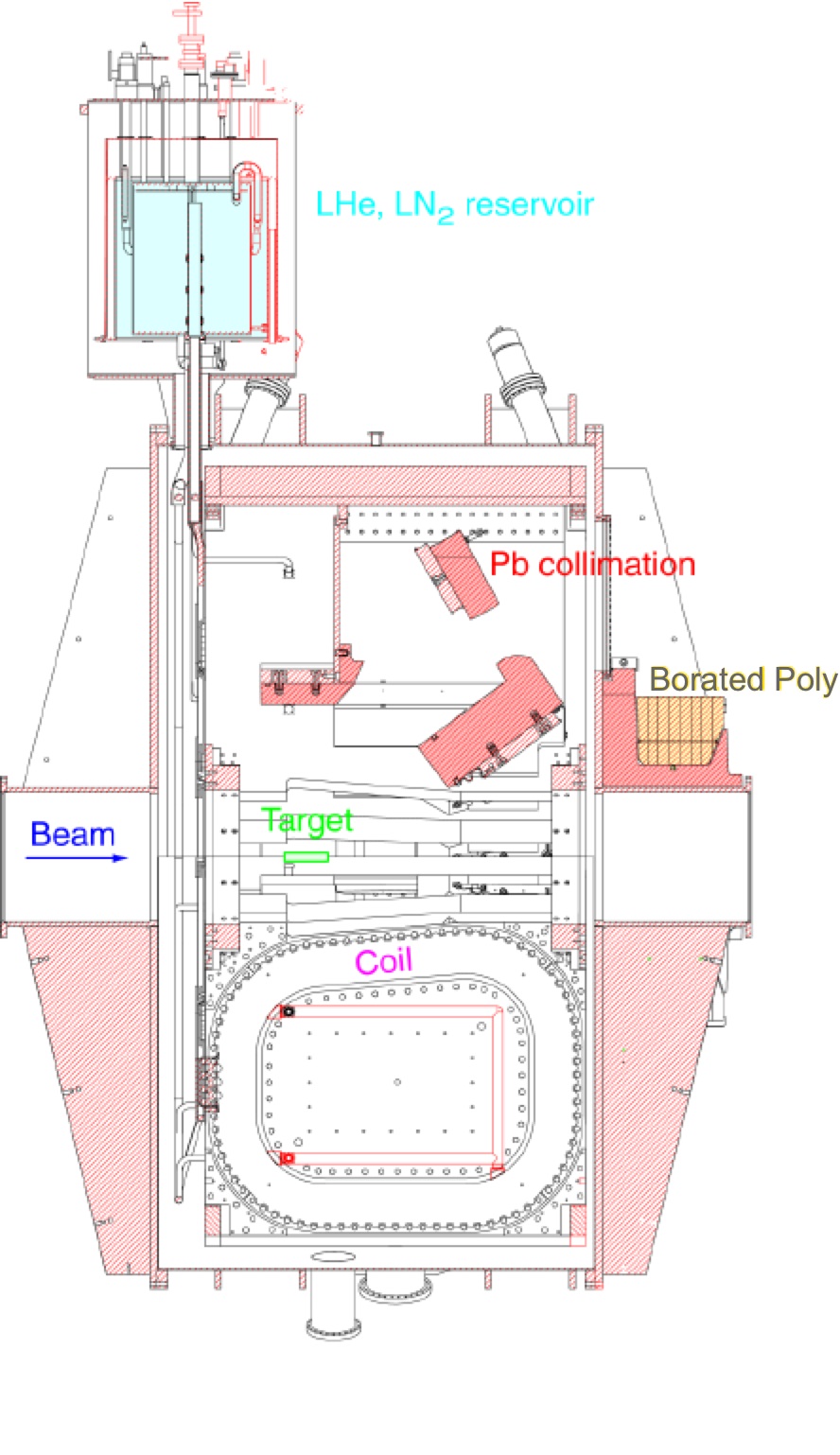}
\end{center}
\caption{Cut-away view of the
interior of the G$0$ SMS, showing the placement of
the target, magnet coils, collimators and shielding, and the
LHe and LN2 reservoirs. In the upper half only the 
collimation is shown and in the lower half only (a section through) the magnet
coil is depicted.  The lead collimators are supported within a
pie-shaped box as discussed in the text.  The U-shaped coil cooling channel is shown within the coil bobbin (outlined by the double lines on the inside of the overall coil package).}
\label{fig:sms-cutaway}
\end{figure}

The collaboration's reference conceptual
design 
was developed by 
the primary contractor for magnet fabrication, BWX Technology,
Inc. (BWXT)~\cite{BWXT}. The magnet was constructed over a period of 3.5 years
at the facilities of BWXT in Lynchburg, Virginia. Control system
development and the initial cooldown were then carried out at the
University of Illinois beginning in 2001. 
The magnet was shipped
early to Jefferson Lab in 2002 for installation in Hall C. In
Fig.~\ref{fig:sms-cutaway}, a vertical cross-section along the
beam line shows some of the principal parts of the spectrometer
which will be discussed in more detail below (see also Ref.~\cite{Brindza}).

\subsubsection{Mechanical design}
\begin{figure}[tbp]
\begin{center}
\includegraphics[width=4in]{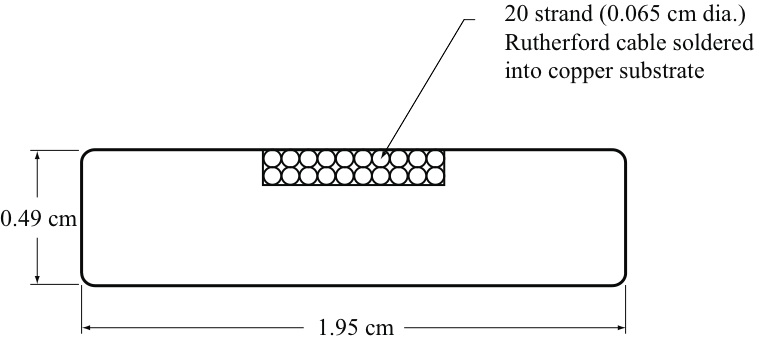}
\end{center}
\caption{Cross-section of superconducting cable.}
\label{fig:sms-conductor}
\end{figure}

The G$0$ spectrometer magnet is made up of eight
superconducting coils configured radially around a central bore
region.  The coils are wound from NbTi superconducting wires in a
Rutherford cable configuration, soldered into a copper substrate
as shown in Fig.~\ref{fig:sms-conductor}. 
The coils are ``dry wound'' (i.e., not ``potted'' with resin) on aluminum coil forms or bobbins in
two ``double pancake'' windings with each layer in a pancake
containing 36 turns, yielding 144 turns total.  The two double
pancakes of a coil are spliced together by overlapping and
soldering the adjacent leads.  Two layers of 0.25~mm G-10 sheet
insulate the pancakes from each other, from the bobbin, and from the
coil case. Each coil is contained in an aluminum case which
provides conductive cooling from the central bobbin and a
mechanism for pre-loading the coils. The Lorentz force between
windings of the coils tends to separate adjacent turns leading to
reduced thermal contact and potentially to movement which can release
sufficient energy to quench the superconductor.  To counter these
 forces and to reduce thermal resistance, the coil windings
are clamped and pre-loaded using pairs of jack-bolts
mounted in blocks attached to the coil side plates.  
Typically the pre-load is 12 kN/cm of
coil perimeter.

\begin{figure}[tbp]
\begin{center}
\includegraphics[width=5.5in]{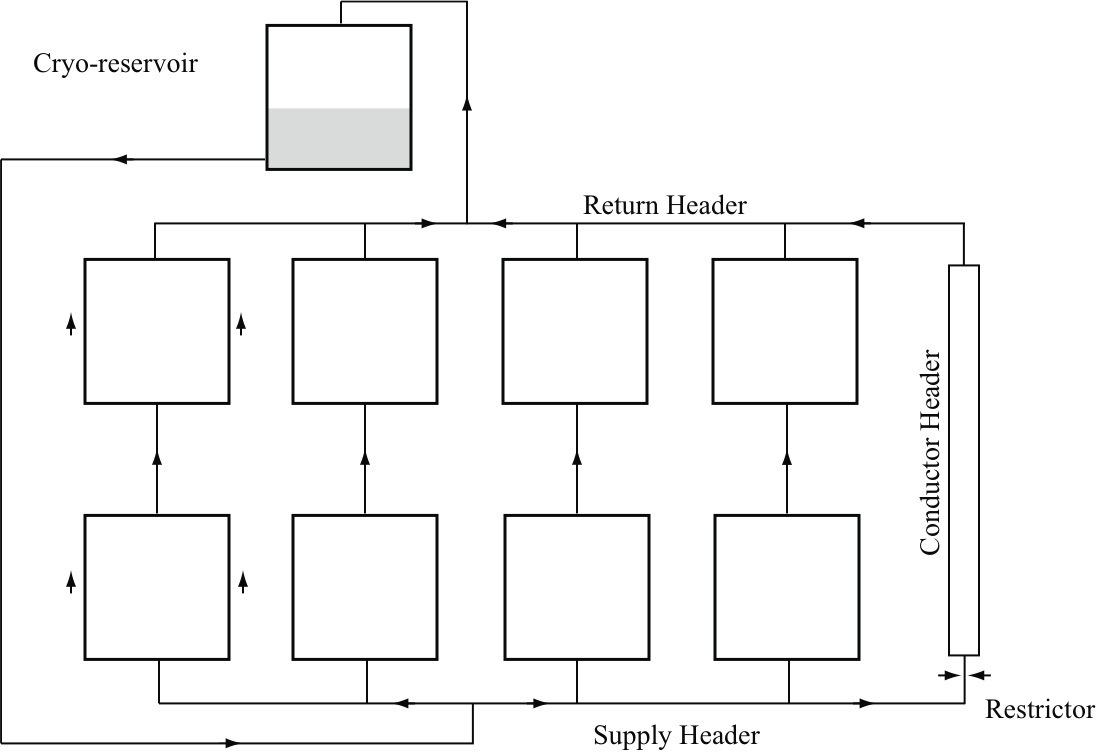}
\end{center}
\caption{Liquid helium thermal siphon circuit for SMS cooling.}
\label{fig:sms-cooling}
\end{figure}

The conductor is cooled to its operating temperature of 4.5~K by
conduction to the central bobbin.  The bobbin, in turn, is cooled
by liquid helium (LHe) passing through an interior U-shaped
channel (see Fig.~\ref{fig:sms-cutaway}).  
Four parallel cooling circuits, each consisting of
two coils and interconnecting plumbing are fed %saturated 
LHe at
the lowest point in each circuit via a supply manifold from a
reservoir at the top of the magnet.  Each circuit is arranged so
that the vertical component of flow is always upward, as shown in
Fig.~\ref{fig:sms-cooling}. As the LHe flowing through the coils
absorbs heat, some boiling occurs, decreasing the bulk density of the
fluid. Density differences between the column of 
LHe in
the supply manifold and the lower density two-phase fluid in the
coils causes the two-phase fluid to rise, carrying the vapor back
to the LHe reservoir.  In the reservoir, the evaporating gas is
returned to the Jefferson Lab ESR, while liquid
is recirculated through supply manifold to the coils. 

The eight coils of the spectrometer magnet are bolted to two
aluminum, load-bearing hubs at their up-stream and down-stream
ends.  
The aluminum hubs are designed to react against the 52 tonne
net inward Lorentz force generated by each coil. The
coils are attached to each other on their outer perimeter by
up-stream and down-stream octagonal aluminum rings. 
The allowed coil positioning uncertainty of $\pm
2$~mm is sufficient to reduce the force between coils due to
misalignment away from symmetry to negligible levels.  

In order to implement the shielding and collimation
discussed in Section~\ref{sec:sms-optics},
a collimator module, containing all the
shielding internal to the cryostat for a single octant, is hung
from the straight sections of the octagonal rings between each
pair of adjacent coils (see Fig.~\ref{fig:sms-cutaway}).  Each module is a roughly pie-shaped box
made of bolted and welded aluminum.  Cast lead-alloy blocks are
fastened to, or suspended between, the two sides of the box which
parallel the coils. A Pb/Ca/Sn/Al alloy~\cite{Pballoy} is chosen
both because of its
mechanical hardness and because it has a lower critical field than pure
lead.  
Differential thermal
contraction during cool-down between the aluminum structure and
lead is significant.  
To ensure the contraction takes place in
a predictable and reproducible way, 
the lead is installed
using a ``spring-beam'' fastening scheme.  Each block of lead is
only bolted to a single plate of the aluminum structure and
fully constrained at only a single point.
Any additional fasteners are mounted to spring beams, machined
out of the aluminum plate, which could flex and therefore allow
relative motion in a less critical direction but not
perpendicular to it.

The cold mass, consisting of the coils, collimator modules, and
aluminum support structure, and with a weight of 
30 tons, is
suspended from the top of the vacuum vessel with four angled tension rods
fastened to the octagon rings at the upper corners. 
The 2.36~cm diameter, 68~cm long rods, made of Inconel 718 
for high strength and low thermal conductivity, are equipped with
ball joints at both ends and a mechanism for fine
length-adjustment at the warm end.  The cold mass must be
restrained axially and laterally, but allowed to move vertically
during cool-down; this constraint is provided by a 3.81~cm
diameter, 316 stainless steel ``shear pin'', which is bolted to
the center 
of the bottom collimator module.
As the cold mass contracts vertically upward during cool down, the
pin slides 
in a G-10 bushing fixed 
to the
vacuum vessel.

Heat due to radiation and conduction is intercepted using a single
liquid nitrogen (LN2) shield which surrounds the entire cold mass including
the central bore (for radiation entering from the beamline).
Thermal siphon flow is used to circulate LN2 from the reservoir at the
top of the cryostat through cooling tubes clamped to the outside cylindrical
surface of the shield.
The shield end-caps and the central bore tube are
cooled by conduction through the aluminum support
structure from the outer cylinder.  Cut-outs in the central tube,
aligned with the apertures between coils, permit scattered
particles to pass from the target into the magnetic field.  No
such provision for particles exiting the magnet is made 
as multiple scattering at that point in the
particle trajectory has a small effect on the position at the
detectors.  The exterior of the shield, as well as the interior of
both ends of the shield, are covered with 20 layers of multi-layer
insulation (MLI).

The cold mass and shield are contained in a single vacuum vessel
consisting of a 4~m diameter, 2~m long cylindrical shell and two
end-caps. The shell is fabricated from 1~cm thick 
low-permeability ($\mu<1.02$) stainless steel reinforced with four
2.5~cm thick stiffening ribs and two 5~cm thick end flanges (see 
Fig.~\ref{fig:sms-cutaway}).
Additional flanged radial penetrations in the cylindrical shell
provide ports for the cold-to-warm supports, instrumentation feedthroughs, pressure relief
safety valves, and vacuum pumps.  Due to the sensitivity of the
magnetic verification scheme (see
Section~\ref{sec:sms-verification}) 
the end-caps are each fabricated from a
circular plate of 3.8~cm thick 6061-T651 aluminum stiffened
against deflection by eight roughly triangular 7.6~cm thick ribs
connecting the circular plate to a 60~cm inner diameter
central beam pipe.  The end-caps are bolted to the flanges on the
ends of the cylindrical part of the vessel with o-ring seals.  
Eight cut-outs in the end-cap nearest the detector
are covered with 0.5~mm thick titanium (Timetal 
15-3~\cite{Timetal}) sheet sealed with o-rings to provide exit windows for
particles passing out of the vacuum on their way to the detectors.

Cryogens 
are supplied 
through a cryo-reservoir
mounted on the top of the vacuum vessel.
This reservoir, patterned after similar devices in Jefferson Lab's Hall C, contains all the valves, controls, and
feed-throughs related to the cryogenic system. 
It consists of an annular 70~l
volume for storage of LN2 surrounding a 160~l cylindrical LHe
reservoir.  Both volumes are surrounded by vacuum gaps and are
wrapped with MLI.  For each of the cryogen circuits,
computer-controlled cryogenic valves allow the selection of either
thermal siphon flow for normal operation or, to expedite
cool-down, direct ``forced flow'' cooling of the magnet.
Control system PID feedback loops adjust the
reservoir supply valves, based on liquid level sensor
measurements, to maintain the cryogen levels in the reservoirs.
The cryo-reservoir also provides the environment for the ``vapor-cooled lead'' feedthroughs
that provide the
transition between the room-temperature, water-cooled power leads
and the superconducting power buss (see
Section~\ref{sec:sms-elec}).

The spectrometer magnet is mounted on an aluminum base frame,
which provides sufficient adjustment to precisely align the
spectrometer to the beam line.  The entire spectrometer, including
the base frame, is mounted on rails which permit the apparatus to
be moved off the beam-line when necessary.

\subsubsection{Auxiliary shielding\label{sec:sms-shielding}}

A major source of
background in both the forward and backward modes is 
radiation showering from the downstream beam line and internal
support hardware. On the end-cap through which the particles exit, 
lead alloy blocks nominally 10~cm thick are installed between the
ribs in the region nearest the central beam tube
to shield the detectors from this
background and to augment the line-of-sight shielding from the
target.  
In forward mode, a supplementary block of 5\% (by
weight) boron loaded polyethylene 
nominally 35~cm thick is
positioned outside these lead blocks
to shield against secondary neutrons produced in the beam line and
lead shielding.  Also in the forward mode, a similar, cylindrical beamline shield of
lead and borated polyethylene extends about 400 cm downstream of the cryostat, 
through the inside hub of the detector support and up to a 130 cm thick wall of iron blocks
used to shield the system from radiation backstreaming from the beam dump.  The beamline
shield is important in reducing detector phototube anode currents to acceptable levels.  For the backward angle measurement, electromagnetic
background produced inside the cryostat is significantly reduced by a cylindrical
shell of lead, 2.5 cm thick, located just inside the central LN2 shield and extending from the
downstream end of the target to the cryostat end-cap.

\subsubsection{Electrical design \label{sec:sms-elec}}

\begin{figure}[tbp]
\begin{center}
\includegraphics[width=4in]{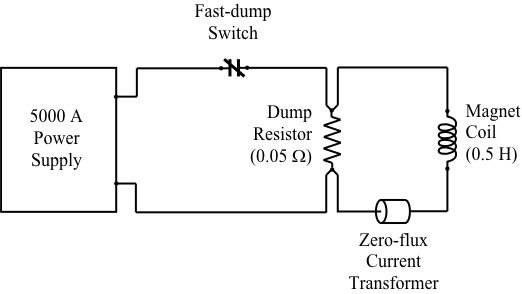}
\end{center}
\caption{Schematic diagram of the electrical circuit of the SMS.}
\label{fig:sms-elec}
\end{figure}

A block diagram of the SMS charging circuit is shown in
Fig.~\ref{fig:sms-elec}. Current is provided by a Dynapower
8000~A, 20 V silicon-controlled-rectifier-based 
supply~\cite{Dynapower}.  A zero-field current transducer is employed to
measure the current supplied to the magnet and to provide feedback
to the power supply for current regulation.  
The power supply architecture supports bi-directional power flow
allowing the magnet to be both charged from, and discharged to, the
 power grid. This feature is used to provide a
``slow-dump'' ($\sim 900$ s) function whereby the magnet current is zeroed by a
powered discharge.  A ``fast-dump'' capability is
provided by a high-speed circuit breaker~\cite{dumpswitch} (rated at 6000 A and 1 kVDC) and an air-cooled high power
resistor~\cite{dumpresistor} (rated at 250 volts, 5000 A with a
resistance of 0.05~$\Omega$). Quench detection, as well as a
number of safety-related faults, initiate
a fast dump. In that event, the fast dump switch opens, disconnecting 
the power supply from the magnet. The
stored energy of the magnet, 6.6~MJ at full excitation, is then
dissipated by the (parallel) dump resistor with a decay time measured to be
10.4~s.   

All interconnections among room-temperature components of the
circuit are made with either 600 (short interconnects) or 1000 MCM (main magnet
connection) flexible, water-cooled
cable. Power connections to the superconducting buss are made via
vapor-cooled leads (VCLs), which pass through the helium
reservoir. They are designed to tolerate the 900~s slow dump after
a complete interruption of coolant flow. Each lead consumes
1.8~l/hr/kA of LHe. The flow of coolant gas is regulated, based on
the magnet current, and measured by flow controllers interfaced to
the control system. The cold (superconducting) electrical bus 
connects each coil in series.  In order to cancel the field from
the cold bus, the current is returned along a path which
parallels the supply path.

\subsection{Tolerances and magnetic field calibration \label{sec:sms-verification}}
Studies 
of the dependence of the SMS forward angle
$Q^2$ resolution on coil profile and alignment, and of the
dependence of magnet symmetry and edge scattering effects on
collimator position generate a number of manufacturing
tolerances for the coils and collimators, 
%some of which are 
listed
in Table~\ref{tab:sms-tol}.

\begin{table}[tbp]
\begin{center}
\begin{tabular}{p{3.5in}l}
Quantity & Value \\ \hline
Deviation of coil current centroid from coplanarity & $\pm 1.0$~mm \\
Deviation in the coil current-centroid from the specified in-plane profile & $\pm 2.0$ mm \\
Error in any direction in the location of a coil in the toroid assembly & $\pm 2.0$ mm \\
Error about any axis in the orientation of a coil in the toroid assembly & $\pm 0.15^\circ$\\
Tolerance on the location of any collimator edge & $\pm 5.0$ mm \\
\end{tabular}
\end{center}
\caption{\label{tab:sms-tol}Manufacturing tolerances for the SMS coils and collimators.}
\end{table}

The current centroid location (flatness and profile) of each coil
was checked prior to assembly by room-temperature magnetic field
measurements. 
The assembled
coil locations were then verified optically with the magnet both
warm and at LN2 temperature at BWXT using digital photogrammetry, made possible
by replacing the Ti exit windows with frames containing optically flat viewports.
At Illinois, the coil locations were
checked magnetically with the SMS cooled to LHe temperature and
running at 20\% of full operating current.
This test,
was
performed using a computer-controlled field mapping system
~\cite{Mapper}, 
which
measured the field components outside the cryostat using
3-axis Hall probes. The
locations of zeros in the field components are determined by
making measurements along selected lines just
outside of the detector end of the cryostat.  The coil locations
and orientations are then inferred by fitting the zero locations
assuming an ideal model for the individual coils. A final digital
photogrammetry check of the coil positions was performed with the
magnet at room temperature after it was delivered and installed at
Jefferson Lab. 
These alignment checks are consistent with the tolerances listed
in Table~\ref{tab:sms-tol}.

Ultimately, the dependence of the G$0$ measurement uncertainty
on magnet tolerances enters through the absolute calibration of
the mean $Q^2$ associated with the focal plane detectors. This
calibration is established in the forward angle mode (and simply transferred to the backward angle measurement)
using the ToF difference between
pions and elastic protons for
each detector. 
A comparison is
made~\cite{Ba04}
between the simulated and measured
ToF differences allowing the magnetic field to vary. 
Simulation and measurement agree to a precision of 100~ps, which
implies an uncertainty on $Q^2$ within the 1\% requirement of the
experiment.  These results are consistent with the uncertainty in the absolute 
magnet current measurement.

\subsection{Control system\label{sec:sms-controls}}

The monitoring and control system for the SMS has 
three principal subsystems:
sensors and signal processing electronics (located in a shielded
location in the experimental hall), a Programmable Logic
Controller (PLC) and its ``ladder logic'' software, and a console
(user interface) computer (located outside the hall in the experiment
data-acquisition area).
Because of radiation-related faults, the PLC was moved from the shielded area
in the experimental hall
to the data-acquisition area between the forward and backward angle run periods, communicating
by ethernet with a ``relay'' module remaining in the shielded area.

The signal processing electronics provide conditioning and interface functions (level
shifting, gain adjusting, isolation, etc.) for the PLC input/output (I/O).
For the most part, the signal
processing electronics are packaged in modular DIN-rail-mounted
components connected directly to the PLC input/output (I/O)
modules.  A notable exception are the resistance thermometers,
read-out using Lakeshore Model 218 temperature monitors~\cite{Lakeshore}.  Serial RS-232
outputs from the monitors are converted by a PC-104 single-board computer
for direct transfer to the PLC memory via ethernet.

The system is controlled by a Direct Logic DL405 PLC~\cite{AutomationDirect}.
The PLC program provides the following
functionality (using, primarily, I/O modules from the manufacturer):
\begin{itemize}
\item analog inputs (cryogen and cold mass temperatures, cryogen pressures, vacuum pressures, valve positions, quench
protection voltages, power supply parameters, etc.) are scaled to ``engineering units'' and
stored;
\item scaled analog values are compared to operator-set levels and alarm indicators are latched
 when a level is exceeded;
\item based on alarm indicators and digital inputs, interlocks are tripped initiating a fast or slow dump of magnet power;
\item cryogenic valve actuators are adjusted according to operator-set values;
\item power supply current is adjusted based on an operator-set current and ramp rate;
\item gas flow through the VCLs is adjusted either based on operator set values or automatically, based on current; and
\item cryogen level and cooldown (or warm-up) are controlled using PID loops.
\end{itemize}
The PLC
executes its ladder logic program with a cycle time of about
25~ms.  However, due to the eight-fold multiplexing of analog
input signals, the minimum guaranteed response time is eight times
longer, 200~ms. 
The PLC also serves as a repository for operator-set parameters
from the control console.  Through a serial connection, all magnet
control data is made available to the lab-wide EPICS control
system~\cite{EPICS}.  This allows the status of the magnet to be monitored by
accelerator and refrigerator operators and provides a
convenient method of including magnet status information in the
experiment's recorded data stream.

The third component of the control system is the console.  It runs a dialect of the National Instruments
human-machine-interface program LookoutDirect~\cite{AutomationDirect} during normal
operation of the magnet. 
This console computer, running Microsoft Windows XP, has three main
functions. First, it serves as a development station (using DirectSoft32) for the
ladder-logic program running on the PLC. 
Second, using screens created with LookoutDirect,
one can monitor the status of analog and digital signals, alter
operator-settable parameters in the PLC, command the PLC to
perform control procedures, and display the logged history of PLC-acquired 
data.  
Finally, the console
performs the function of logging all monitored data and control
parameters to a local, mirrored, hard disk.
Data are stored as a
function of time in compressed form by LookoutDirect and are
available to any Microsoft Open Database Connectivity (ODBC)
compatible application.

Quench detection is provided by two parallel systems, a digital
PLC-based system and a hard-wired analog system.  Both systems are
designed to be insensitive to the inductive voltages produced during
ramping, and to provide detection of quenches, not only in the
coils, but also in the superconducting buss leads. The digital
quench protection system is considered a ``backup'' for the analog
system.

Three voltage taps per coil (two on the coil leads and one at the
interconnection point between double pancakes), provide a measure
of the coil voltage for the quench-detection systems.  In
addition, voltage taps are located  on the ``transition leads'' on
the cold side of the VCLs within the LHe reservoir, and about
midway between the VCLs and the first coil connection. Finally a
diagnostic voltage tap is made at the halfway point on the long
return of the cold buss after the last coil.

The analog quench detection system follows a design developed for
the CDF~\cite{Dr99} and D0~\cite{Ha98} experiments at Fermilab.
The magnet forms one half of a bridge circuit.  A resistor chain
spanning the power supply connections provides the other half. 
A single quench
within the magnet will unbalance the bridge.  Isolation
amplifier/window discriminator channels sense the imbalance at a
threshold of about 350 mV 
and
trigger a fast dump.  Additional isolation amplifier/discriminator
channels monitor voltage taps across the supply and return
superconducting leads, comparing voltage drops to current-dependent 
thresholds.

Employing isolation amplifiers, the digital quench detection system 
uses the PLC to examine voltage
drops between taps.  To accommodate different operating conditions,
the PLC compares
each voltage to an average voltage derived from all of the coils.
A
deviation greater than a set threshold of around 200 mV generates
a
fast dump. 
Separate thresholds are defined for the
transition lead voltage drops, which are also monitored by the digital quench
detection system. 

\subsection{SMS performance\label{sec:sms-performance}}

Since its installation at Jefferson Lab, the G$0$ SMS has been
operated at full power (5000 A) during several runs extending over many months (forward mode)
and then for a period of over a year at lower currents (3500 A and 2650 A - backward mode)
With only a few exceptions, noted below, it has
performed according to the specifications of the original design.

In the reference design, the heat load to LHe is specified to be less than 40~W.  However,
boil-off studies indicate that the as-built load is about 107~W.  
The
steady-state LHe requirement of the magnet at full power is
measured at Jefferson lab to be about 8~g/s.  This is consistent
with the measured heat load and some additional load from the
supply lines.

Magnet cooldown to LN2 temperature is accomplished by regulating
the temperature of helium gas flowing in the cooling circuit
through the use of an LN2 heat-exchanger.  This part of the
cooldown requires about 17 days, limited by the cooling capacity
of the heat-exchanger and 
by the safety requirement
that $\Delta T$ between inlet and coil
average be less than 75~K.  Overall, the cooldown typically extended over 21
days, about the twice the original specification.

About 160 of the 3270 hours ($\sim 5$\%) of available data collection time
during the forward mode commissioning and production running were lost
because of problems with the SMS. This represents about 50\% of the
lost data collection time (the rest was lost due to problems with
other systems: target, DAQ, etc.). Most (70\%) of the magnet problems
were caused by radiation-related faults in control system components located
in the experimental hall.  The most common such problem was the result
of non-permanent, radiation-related changes to PLC software
that triggered fail-safe logic resulting in
a fast dump of the magnet power. 
Once the PLC program was
restored and executing, a minimum of 2.5 hours was required to recover
LHe level and restore the magnet current.  Most of the rest of the magnet-related
downtime in the forward measurement was associated with failures of relays
used to power the valve motors in the cryogen level control system.

After the PLC was moved to the experiment data-acquisition
area, communicating with a relay module interface to the I/O electronics via
ethernet, the system was more reliable, because the relay module was much more radiation resistant.  The valve control
relays were also replaced with a solid-state switching system.
With this new configuration, magnet controls problems accounted for about a 
1\% loss of data collection time.

\section{Particle Detectors}                                                   
\label{sec:detectors}
The G0 particle detection system is composed of eight octants
of detectors, that count recoil protons from small angle
($7-13^\circ$) e-p scattering (forward angle mode, initial
orientation of the magnetic spectrometer), and scattered electrons
from large angle ($110^\circ$) e-p scattering (backward angle
mode, second phase of G0 following a $180^\circ$ rotation of the
spectrometer).

\begin{figure}[tbp]
\begin{center}
\includegraphics[width=4in]{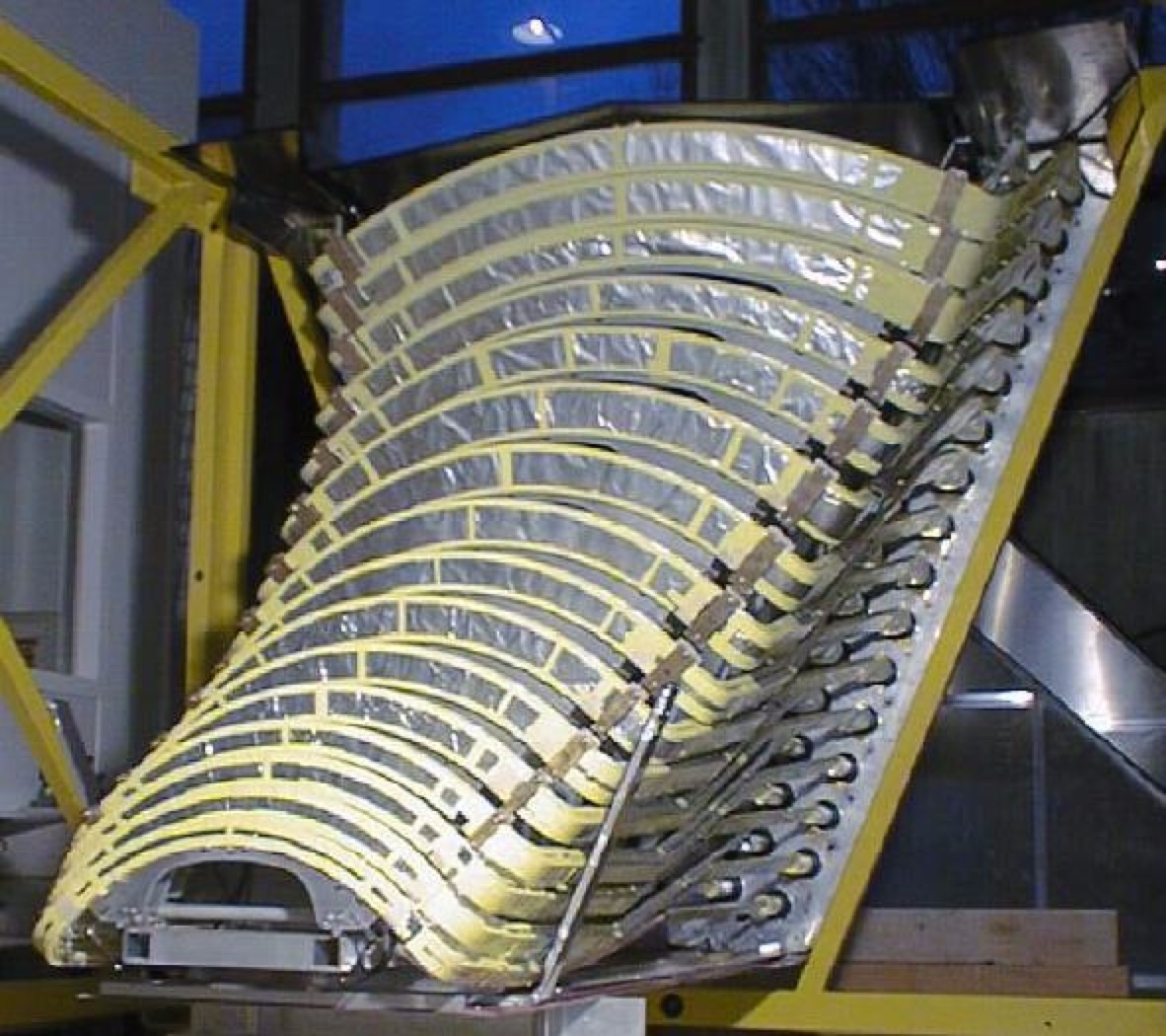}
\end{center}
\caption{Photograph of one FR FPD octant. The FPD
numbers (1 to 16) increase from the bottom to the top.  The scintillators
are the arc-shaped segments lying in (nearly) vertical planes.  Light guides from the ends of the arcs lead to phototubes at the back plane of the mounting
structure.}
\label{fig:FPD}
\end{figure}

For the forward angle measurements, there are 16 detectors in each
octant lying on, or near, the focal surface of the SMS magnet.
These FPDs detectors
(see Fig.~\ref{fig:FPD}) consist of pairs of plastic
scintillators, contoured to define a specific range of $Q^2$. The
first 14 detector pairs each measure a rather narrow range of
$Q^2$ values of the recoil proton for e-p elastic scattering in
the domain 0.12 to 0.55~(GeV/c)$^2$. As mentioned in Section~\ref{ss:forwarddesign},
most of the background from inelastic protons and pions in these
detectors is eliminated by measuring the ToF of
the particles to the detectors. By contrast, detector 15 collects
recoil protons over the $Q^2$ domain from 0.55 up to approximately
0.9~(GeV/c)$^2$. For this detector the ToF serves as a
measurement of the momentum of the proton, and thus the $Q^2$.
Because of the ``turnaround''
of the higher 
energy protons (Section~\ref{ss:forwarddesign}), the yield
for elastically scattered protons with a $Q^2$ near
1.0~(GeV/c)$^2$ actually falls on FPD 14, but at a very different
location in the time spectrum. Finally, the $16^{\rm th}$ detector
serves as a monitor of backgrounds and the
spectrometer field.

For the backward angle electron scattering, the kinematics are such
that a single $Q^2$ value is measured for each incident beam
energy. In the case of the backward angle measurements,
ToF is no longer useful in terms of separating the
elastic, inelastic and pion events, since the flight time is
essentially the same for these relativistic particles. Instead, a second array of nine scintillators is added near the exit
window of the magnet cryostat for each octant. As discussed in
Section~\ref{ss:backwarddesign}, these CEDs are used in coincidence with 14 of the 16 FPDs
in order to define the trajectory of the particles. Using different sets of CED-FPD pairs, electrons from elastic e-p
scattering can be separated from those associated with inelastic
scattering. Simulations and analysis of test data~\cite{Tieulent2}
indicated the need to add 
a \v Cerenkov detector 
to each octant to discriminate electrons from pions. This particle
identification is crucial for the measurements with the deuterium
target, since there is a large yield from negative pions
produced from interactions with the neutrons.

\subsection{Focal plane detectors}

The FPD scintillators are arch-like shapes which are defined
by tracing an array of proton rays, from elastic e-p scattering,
through the SMS magnetic field using TOSCA~\cite{vectorfields} and  
specialized fast tracking codes.
Specifically, based on
simulation, elastic e-p scattering Q$^2$ bins are chosen to
provide reasonable resolution and to produce roughly equal count
rates, at least up to about FPD number 11, above which the e-p
elastic rates are reduced and the detector widths are chosen on the basis of
momentum resolution. 
Rays are generated for three positions along the 20~cm cryotarget (front, middle, and rear) and for three azimuthal
angles $\phi = 0.0, 5.0,$ and $10.5^\circ$ (the rays are
symmetric about the center of the octant, $\phi = 0^\circ$). The
azimuthal acceptance for each octant is limited to $\phi=\pm
10^\circ$ by the upstream collimators in the SMS magnet, and
$0.5^\circ$ is added to account for possible misalignment of
these collimators. To design each of the first 13 FPDs, the
rays corresponding to the lower and upper boundaries of the Q$^2$
bin are projected onto an x-y plane at the z location of the
focal surface for that Q$^2$ and the central ray $\phi = 0^\circ$.
For the central rays, the x-y positions for the three target
positions coincided, as expected for the SMS design. However, for
non-zero $\phi$ angles the positions differ
by a few mm due to changes in the magnetic field near the coils.
After examining various descriptions for the shapes of the
detector boundaries and their effect on the Q$^2$ resolution, we
concluded that detector boundaries in the x-y plane are
satisfactorily described by arcs of circles with the center and
radius chosen from a fit to rays corresponding to the three
azimuthal angles and originating from the center of the target. The lengths
(azimuthal extent) of the counters are then determined from the
most extreme rays, which originate from the upstream end of the
target at $\phi = 10.5^\circ$, to which approximately 1~cm in
length is added to compensate for multiple scattering and
possible misalignment (2~mm) of the detectors. Typical dimensions
of the scintillators are 60$-$120~cm in length and 5$-$10~cm in
width. Finally, the defined detector shape is rotated by a few
degrees about the y-axis so particles enter perpendicular to
the detector, at least at its center. This is done to reduce
double hits on the FPDs. For detectors FPD 14 and 15 the same
basic procedure is followed; but 
as discussed above, the one-to-one correlation
between focal plane position and Q$^2$ disappears at larger Q$^2$.
There is a turnaround of the ToF versus Q$^2$ elastic
locus as the momentum becomes relatively larger and the recoil
angle smaller, and both detectors count protons from a range
of Q$^2$ values of the recoiling protons. Note that FPD 16 is
chosen to have a shape identical to that of FPD 15, since it serves as a monitor for
background. The final detector design is done with TOSCA-based
software, 
the end result being a file for computer-controlled machining.

Each of the FPD scintillators is paired with a second identically shaped
partner to reduce background from neutral and low energy particles. In the interest of
redundancy and time resolution, for the forward angle measurement, the two ends of each scintillator are viewed
with a photomultiplier tube (four PMTs per scintillator pair).  With less strict timing requirements, the backward angle measurements utilized the same scintillator pair, but with only one PMT on each member (e.g., at the left end of the front detector and the right end of the back detector).  Lucite light
guides are used to transmit the light produced in the scintillator to
photomultiplier tubes mounted in a region of low magnetic field - as much
as 2 m away for the low Q$^2$ FPDs (see Fig.~\ref{fig:FPD}). The high density of light guide material
on the sides of each detector pair required significant design time to
overcome interferences between the various components.  The length and thinness of the scintillators and light guides led to
concerns about the number of photoelectrons from the photomultiplier
tubes, but a set of simulations and experimental studies were undertaken to
confirm that a satisfactory number of photoelectrons would be produced by
minimum ionizing particles. 

Internal alignment to 2~mm of the 16 scintillator elements making up
a single octant is carried out when the octant modules are assembled.
The alignment of FPD octant modules relative to the magnet and electron beam
is accomplished with adjustment degrees of freedom provided by the detector
octant support frame.

For several reasons, based on budgetary constraints and on
technical and scientific grounds, the detectors and electronics
for the four octants numbered
1,3,5, and 7 were built by a North-American (NA) collaboration
(USA-Canada), and those for octants 2, 4, 6, and 8 were built by a French
(FR) collaboration.
Although the basic elements of the detector systems are identical,
there are differences in the details of the design and the
construction procedures.
 The important differences are described in the following two
subsections.  Note that the assigned octants for each collaboration 
are opposite each other
in azimuth, in order to reduce possible systematic
errors.

\subsubsection{North American FPD Detectors}

The NA scintillation detectors are made from Bicron BC-408~\cite{Bicron}. The
FPD~5-16 pairs are fabricated from 1~cm thick scintillator
sheets, whereas FPD~1-3 are fabricated from 5~mm scintillator to accomodate
the lowest energy protons. FPD~4 consists of a 0.5~cm front layer
followed by a 1~cm rear layer of scintillator. 
The scintillator pairs are separated by identically-shaped
3 mm thick polycarbonate sheets to provide additional absorption of low-energy particles. All scintillators
are initially rough machined to the approximate shape (1.5~mm
oversized)using water jet cutting. Five of these scintillators are then
stacked, and the curved sides milled on a CNC machine. The
machined sides are hand polished. The quality of the
polished surface for each detector is tested using an automated
laser reflection technique, and several were re-polished to improve performance. In the final step each scintillator is wrapped
using strips of aluminized mylar. It is not necessary to make
this wrapping light-tight, since the octant support is designed
as a light-tight box.

The NA light guides are fabricated from UVT transmitting lucite
(Bicron BC-800~\cite{Bicron}). Due
to the complicated geometry of the light guides, it was necessary to
develop a series of jigs and bending techniques. Silicone
``cookies'' are used for coupling the light guides to the
phototubes.

Once the scintillators and light guides are fabricated, they
are mounted and glued on the final precision support structure at Jefferson Lab. After assembly, careful tests of
the performance of each detector are made.
Specifically, a $^{106}$Ru $\beta$ source is placed at several
locations along the length of the scintillator and the light
output measured. The design goal is to obtain $\geq$100
photoelectrons for proton detection and $\geq$50 photoelectrons
for electron detection for the worst case, i.e., when the source
was on the far end of the scintillator away from the phototube. In
all cases this goal is exceeded by about a factor of
two. In addition to the source tests, prior to sealing the light
tight box, data were taken for cosmic rays, permitting one to
monitor possible deterioration of the scintillators, light guides,
and/or photomultiplier tubes over time.

The phototubes for the four North American octants are
12-stage Photonis~\cite{Photonis} XP2262B tubes.  All tubes were tested for
non-linearity and their gains measured with the results stored in the experiment database. 
The design of the PMT base takes into account the relatively
large dynamical range required by the experiment (light output
from minimum ionizing electrons to that of 60~MeV protons). The
bases are passive, being made up of resistors and Zener diodes,
and include a Zener-assisted front stage to maintain collection
efficiency of the primary photoelectrons, independent of the
operational setting of the photomultiplier tube. 
The photomulipliers
are also equipped with $\mu$-metal magnetic shielding.

To transport the signals from the PMTs in the experimental area to
the electronics, approximately 150~m of coaxial cable is used.
This leads to significant attenuation in the signals, which in turn
necessitates either operating the tubes at a higher voltage
(leading to more anode current), or introducing an extra stage of
amplification. Due to concerns about high anode currents reducing
the lifetime of the PMTs and the long running time needed to carry
out the G0 experiment, amplifiers with gains of approximately 25 (Phillips
model 776, modified to increase gain from 10 to 25) were introduced in
the experimental area. This addition maintained the anode currents in
an acceptable range.

\subsubsection{French FPD Detectors}

The FR detectors are similar in many respects to the NA detectors; only key differences are highlighted here.  They are also fabricated from BC-408, in this case obtained from Eurisys, a
European subsidiary of Bicron~\cite{Bicron}.  The FPDs 1-3 are 5~mm thick
 and FPDs 4-16 10~mm thick. 
The scintillator pairs are separated by identically-shaped
3 mm thick aluminum sheets. 
The light
guides, made of acrylic, were machined by a contractor in straight sections and bent directly onto the mechanical support
structure. Additional parts of the light transmitting system (fish
tails and PMT adaptors) were also made commercially and later glued to
the light guides.  The most robust joints were glued and the scintillators
and light guides were then wrapped with aluminum foil.

Before the material was shipped to Jefferson Lab, a complete test assembly
of one octant was done in France, in order to
check the procedure and to identify possible interferences. Using cosmic rays, photon yield measurements are made at three locations
along the scintillators. The absolute normalization is done using
the single photoelectron signal from a photodiode.
The French FPD detectors generate essentially the same number of photoelectrons as
compared to the NA design. The final operation, the
gluing of the more delicate joints, took place at Jefferson Lab.
Optical grease was used to couple the PMTs and light guides.

For the FR FPDs, lower gain,
8-dynode Photonis XP2282B tubes~\cite{Photonis} are used and specified to have small gain
dispersion.
This feature simplifies the gain adjustment through
variation of the high voltage. After tests, it is indeed found
that the gain variations measured are less than a factor of 4.
%All the characteristics of the PMTs are recorded in a
%database.

For the photomuliplier bases, a built-in amplifier (gain of 20
overall) is included in the design. The anode currents are thus
kept to values of a few $\mu$A without additional amplifiers, 
allowing the long term operation of the tubes needed by the G0
project. Following irradiation tests of components (see next
section), Zener diodes were chosen instead of transistors for
gain stabilization in the electrical design.
A base-line restoration function is also included in the design. In addition to
the $\mu$-metal magnetic shielding, an electromagnetic shield
consisting of a metallic copper sheet rolled around the plastic
housing is incorporated to decrease the noise on the PMT signal.

The octant support is a welded aluminum tube structure, designed to align the detectors to within 2 mm of their optimal locations. In
order to save weight, the backplane is made of two 20~mm thick
beams that form a V-shape. Finite element analysis was carried out
to ensure acceptable deformations of the support structure.
Finally, a dedicated aluminum structure holding a Tedlar cover
is integrated in the design to ensure the octant is light-tight.

\subsection{Radiation damage tests}

Prior to construction, extensive neutron irradiation tests of the
light guide material and of different types of glue were
performed at a dedicated facility in Orleans (France).  Two 5~mm-thick pieces of acrylic were glued together
and irradiated by a flux of $5 \times 10^{13}$ to
$10^{14}\;\mbox{\rm n}/\mbox{\rm cm}^2$ 6~MeV neutrons. The
electronic components for the French bases and the silicone cookies
used to couple the North American PMTs were also tested.  The results
showed that, after irradiation:

\begin{itemize}
\item attenuation of light by the glue and the silicone cookies 
varies from negligible up to 5\%;

\item increased attenuation of light in the 
acrylic could reach 20\% around 400~nm for large neutron doses (with the implication that during the time between G0 runs when other high-intensity experiments are performed in Hall C, the detectors must be properly shielded); and

\item the Zener diodes, and the base-line restoration and amplifier transistors used in the FR bases suffered no degradation
(total of $7 \times 10^{14}\;\mbox{\rm n}/\mbox{\rm cm}^2$).

\end{itemize}

\subsection{Anode current measurements and PMT protection circuit}

Particularly for the forward angle measurement, there was concern about background events in the
detector PMTs with amplitudes below the discriminator threshold, but
which contribute significantly to the photomultiplier
tube anode current.  This anode current should
be kept below about 50 $\mu$A, with an absolute limit of 200 $\mu$A, in
order to limit gain changes due to PMT aging during the experiment.
Under normal operation about 90\% of the anode current is due to
sub-threshold events, mainly due to GHz-rate single-photoelectron
events. This high background rate results in a nearly DC
background anode current, requiring
reduction of the PMT gain and use of additional amplifiers of gain 20
(FR detectors) and 25 (NA detectors) so as to obtain the necessary
amplitude for proton signals at the discriminators.

To continuously monitor the anode currents, we use the average signal
height as measured in the monitor ADC spectra (see Section~\ref{monitoring}),
including events below the threshold of the main, time-encoding electronics. During data-taking, a part of the acquisition code determines
the average of each ADC spectrum after a user-selected number of
events, subtracts the results obtained without beam, applies a calibration
factor and produces a bar graph of the anode currents for all
detectors. The algorithm can average backwards in time with a
user-selected decay constant to smooth out fluctuations.

The calibration factor is determined by measuring the anode
current for selected tubes with a pico-ammeter (a) directly at the PMT
and (b) after the amplifier. To obtain the amplified anode current and to
correct for the amplifier DC offset, the PMT high voltage is cycled
on and off.

Under certain conditions, the anode currents can increase
dramatically, for example, when diagnostic devices are inserted into
the beam line or when the SMS current is zero. In order to
protect the PMTs from damage under these circumstances, an electronic
``high-voltage shutdown" circuit is included in the detector system.  It has 
analog inputs from 8 selected PMTs, as well as logic signals from the
SMS and M{\o}ller polarimeter magnets.  The analog inputs are integrated with a time constant of 0.5 s
(which approximates the time constant for PMT damage) and with a
threshold at approximately twice the desired maximum anode current.

In addition, the anode current of a test, ``paddle" scintillation detector, 
is monitored. Placed near the main detectors, it serves to continuously indicate the general beam quality, especially when the main detectors are turned off.

\subsection{Cryostat exit detectors}
\label{subsec:CED}

As noted above, since ToF is no longer useful at the
backward angles, the CEDs are required for to separated elastic and inelastic events in
the G0 backward angle phase (see, for example, Fig.~{fig:backwardPhoto}). The array of nine CEDs in each octant is used in
coincidence with the top 14 of the 16 FPDs.  The CED/FPD
combination correlates the momentum and scattering angle of the
detected electrons, and
thus allows for the separation of elastic and inelastic events.
With front-end electronics composed of gate arrays, it is possible
to record events for each CED/FPD pair, and
thereby measure asymmetries for both elastic and inelastic events.

Constrained by the general shapes of the FPD scintillators, the
CEDs are also arch-shaped detectors. They are 1~cm thick BC-408 scintillators coupled to BC-800
UVT light guides~\cite{Bicron}, and are viewed from each end by the same photomultiplier
tubes as used for the FPDs. The CED light guides have shapes similar
to the FPD light guides, primarily involving a tight
"helical bend" which guides the light from a high-field,
geometrically-constrained region, to a lower-field region
approximately 1.5~m away where the photomultiplier tubes are
located. 

The photomultiplier tube arrangement for the CEDs is similar to
that used for the FPDs. In fact, because of the extra detector in the coincidence (the CED), we were able to remove one phototube from each element of the FPD pair and reuse them for the CEDs.  A
detailed simulation of expected light yield from the CEDs
was done and tests performed on prototypes. The expected number of
photoelectrons produced was found to be greater than 50 for
minimum ionizing electrons for each of the two PMTs.  Tests using cosmic muons, 
performed after assembly at Jefferson Lab, showed that these numbers
were conservative.

The design of the octant support structure for the CEDs 
takes into account both the required
mechanical support of the CED scintillator/light guide/PMT and
base assemblies, as well as the relatively weak alignment
constraints on these detectors. 
Because of their physical proximity, an integrated design for the CED-\v
Cerenkov support subsystems is used. The support structure
centers around the use of prefabricated aluminum extrusions from
Bosch~\cite{Bosch} because of their strength, versatility, and
relatively low cost. A series of detailed finite-element analysis
studies was carried out 
to
identify potential problems and to optimize the strength and cost
of the support structure. 

\subsection{\v Cerenkov detectors}

An aerogel \v Cerenkov detector is used to reject $ \pi ^{-}$ background from $n(e, \pi)$ reactions and is especially important
for the backward angle quasi-elastic scattering of electrons from the deuterium target.  The detector operates in coincidence mode for the
electron detection, and in veto mode for background studies
and pion measurements.
It is designed to reject pions over the full G0 momentum range, i.e., beyond a minimum of
$\sim 400$ MeV/c.
% corresponding to $Q^{2}=0.8$ GeV$^2$. 
Eight such detectors were constructed and
mounted between the CEDs and FPDs of each octant.

\begin{figure}[tbp]
\begin{center}
\includegraphics[width=4in]{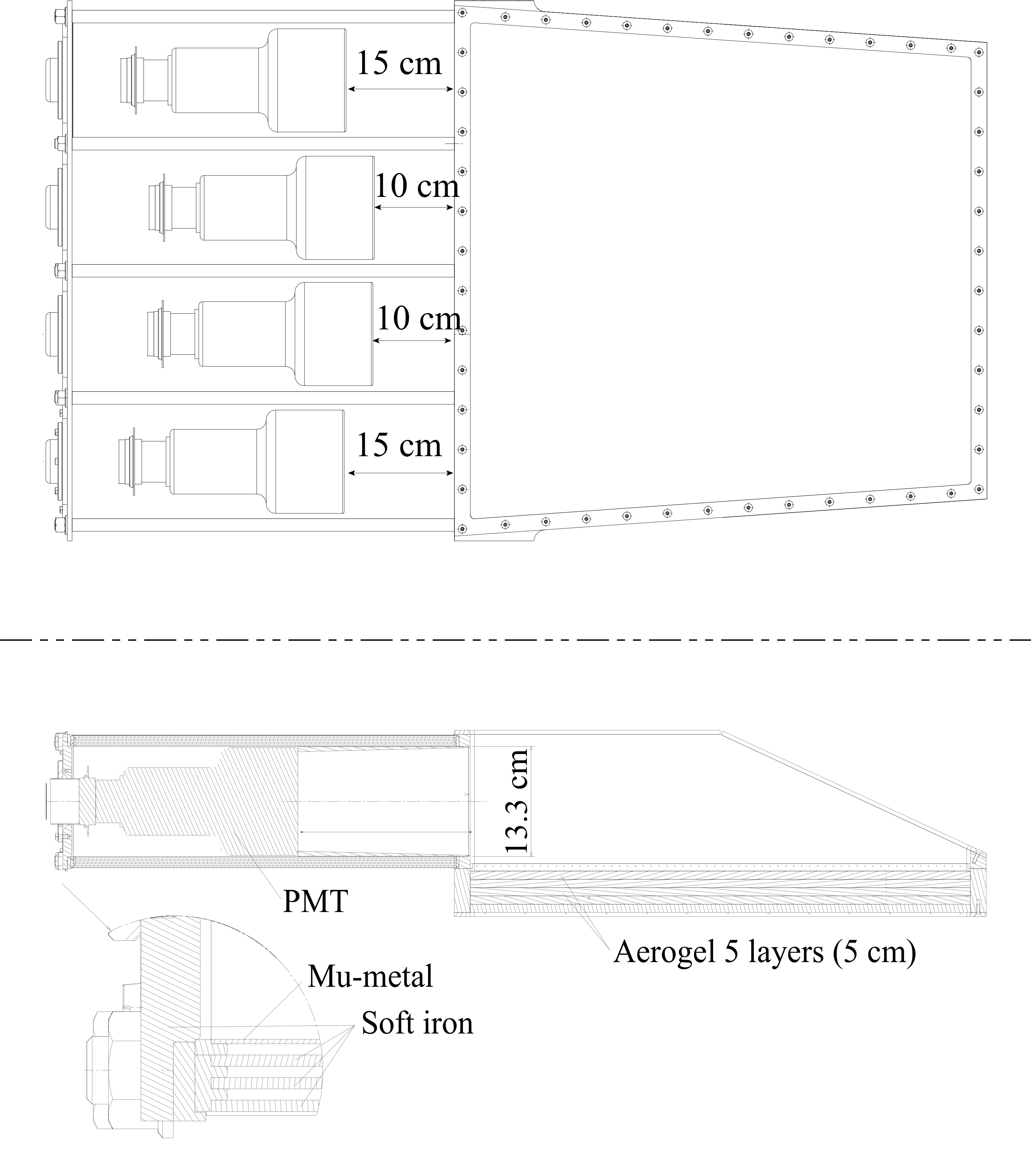}
\end{center}
\caption{Schematic of a \v Cerenkov counter -- beam view and section.  Both the magnetic shielding and the varying phototube set-back distances are necessary to cope with the fringe field from the SMS.}
\label{fig:Cerenkov}
\end{figure}

The electrons and pions pass through 5.5 cm of clear SP30 aerogel~\cite{Matsushita}. The aerogel has 
an index of
refraction of $ n \, = \, 1.035$.  With this index, all primary electrons produce light, but pions below about
570 MeV/c do not. The light is emitted within
a small angle cone ($\sim 15^\circ$)
and enters a downstream region whose walls are lined with a white
diffuse reflector, GSWP 00010 paper~\cite{Millipore}.
This light is collected by four XP4572B phototubes from
Photonis
~\cite{Photonis}
(see Fig.~\ref{fig:Cerenkov}).
The fraction of the light collected by the phototubes is about
 4~\% of that produced in the aerogel. The other goals in the
design are to cover as large a fraction as possible of the
G0 acceptance, while keeping the timing spread as narrow as
possible and the collection efficiency of the photons by the tubes
independent of the position of the particles to be identified.
Studies with both Monte Carlo simulation~\cite{HIG98} and prototypes of the 
\v Cerenkov counter were carried out in France 
and in North-America. 
Prototypes were tested
with cosmic muons 
and 
using test beams at TRIUMF.  A mixed particle test beam was used to measure
the pion-electron discrimination, and the
M11 $\beta \sim 1$ electron beam line was used to measure the position dependence of the overall measurement efficiency.  Electrons are found to generate a signal of 
6-7~photoelectrons. With such numbers, the detection efficiency
for an electron is close to 95\%~, whereas a 400 MeV/c pion has a rejection factor of 125 to 1. The pion ``signal'' is, in fact,
mostly due to $\delta $-rays produced in the CEDs and the \v Cerenkov structure.  

The four phototube signals in each counter octant are
combined together, typically with a 2/4 coincidence requirement.  This signal is
discriminated and used in the trigger.  The typical time-width of the signal is
$\sim$20~ns (due mainly to collection time in the light box), and the
rise time of the pulse is $\sim 1$~ns.  ADC spectra for the individual tubes are recorded by the monitor electronics and used to check the calibration and pion contamination factor of each octant.

Magnetic shielding is important for the
operation of the phototubes in the \v Cerenkov detectors because their sensitivity is reduced by
a factor two when operated in a region with a field as
low as 0.1~mT.  During the G0 experiment, these tubes encounter
maximum field components of the order of 4~mT in the axial
direction and about 11~mT in the transverse direction. Such relatively high
fields require efficient shielding. Tests under similar field conditions were
thus performed at the Grenoble High Magnetic Field Laboratory.
The configuration chosen (see Fig.~\ref{fig:Cerenkov}) was
tested and shown to achieve the required shielding under
the G0 running conditions. 
It consists of three 45 cm long concentric tubes of 0.3 cm thick soft iron and one of 0.5 mm thick Mu-metal (permeability of about $8\times10^4$)surrounding the photomultiplier tube and capped by a square soft
iron plate at the socket end of the tube. 
It is necessary to isolate the
Mu-metal tube from the square plate in order to
break the magnetic circuit.  With this shielding, 
no significant loss 
in gain 
is measured for the 
nominal G0 magnetic field. In addition, these
tests also permitted us to optimize the position of the PMT relative to the counter body and  
adopt 
a set-back of only 10-15~cm. 
Compared to the initial design with a 20~cm set-back, this provides
a significant (of order 30\%) increase of the total number of 
photoelectrons collected by the tubes.

The original borosilicate photomultiplier tubes in the \v Cerenkov detector were eventually replaced with
quartz-faced tubes because of neutron backgrounds (especially prominent in the
deuterium running).  Low energy neutrons capture on the $^{10}$B in the tube faces and sides, producing $\alpha$ particles whose scintillation is detected by the tube.  Using tubes with quartz faces (special order, similar to Photonis XP4572B~\cite{Photonis}) reduced the background signal rate by slightly more than a factor of two. 

\subsection{Gain monitoring system}\label{GMS}

In order to track relative variations in the pulse-height
and time response of the scintillator-based detector system, a
gain-monitoring system (GMS) is used. The underlying
design of the system is similar to those found in many other
experiments~\cite{TAN02}:  a light source generates a short 
burst of photons
that are distributed to the scintillator elements via optical
fibers. Originally, the system was based on a fast nitrogen laser
(pulse length $<$1~ns), but the reliability of these lasers did
not prove sufficient for our needs, and the very short pulses
proved unnecessary, so in the final implementation a nitrogen flashlamp 
is used. The distribution of
wavelengths produced by the flashlamp has a maximum intensity in
the 
ultraviolet, near 350~nm. A tail extends through the visible;
the intensity at 800~nm is about 15\% of that at 350~nm.
Wavelengths below 200~nm are cut off by the exit window of the
lamp. The light from this lamp encounters a rotating mask that
permits the light to fall upon 1 of 15~clusters of optical fibers.
The mask rotates continuously; a set of switches indicates the
location of the mask, and associated electronics control a ``mask
ready'' signal used to make sure the flashlamp will only be fired
when the mask is in a valid position. Each of the 15 clusters
behind the mask contains 19~optical fibers. The fibers are arranged so that, for example,
one of either the ``left'' or the ``right'' end of each FPD
scintillator element is fired at a time. The comparison of the
response of the scintillator and phototubes to these left
and right pulses is used to monitor changes in the condition
of the scintillator separately from changes in the individual
phototube gains.

The flash lamp intensity varies significantly from one pulse to
the next. An additional cluster of 7~fibers is therefore positioned
behind a hole at the center of the mask and
illuminated on every firing of the lamp. Some of these fibers are
connected to small scintillators attached to the window surfaces of
phototubes, whose signals are used to correct for the
pulse-to-pulse behavior of the lamp. These phototubes are
protected from the large flux of charged particles present in the
experimental hall. One scintillator in this referencing subsystem
contains approximately 0.6~nCi of $^{241}$Am. The $\alpha$-decay
from the source produces light pulses used to monitor the drift in the gain of this reference detector.

The optical fibers used are pure silica
 (core and cladding), 
allowing for the transmission of ultraviolet light over a long
distance, and making them resistant to the very high
radiation environment in the experimental hall. By transmitting
ultraviolet light directly to the scintillator, one takes
advantage of the resulting conversion into blue light via the
fluorescence effect which is very similar to the process of
scintillation. This provides a uniform illumination of the detector
volume in a way that is similar to that from the passage of a charged
particle. 

The mask is rotated
so mask ready signals are produced at a rate of about
2~Hz. The mask is ``ready'' for a period of about 70~ms each time.
The actual firing of the flashlamp is controlled by the data
acquisition system which, upon detecting the presence of a new
``mask ready" signal, waits until the next helicity-flip period
(occurring approximately every 33 ms and lasting only 
500~$\mu$s)
to fire the lamp.  Thus, the flashing of the lamp and subsequent
collection of GMS pulse-height and timing data occurs during a
time when no asymmetry data are being collected.

The GMS provides only a relative gain reference; its final
configuration (nitrogen gas pressure, plasma discharge
voltage, and trigger timing) was determined along with the
configuration of the rest of the detector system during our
engineering runs.  After the relationship between the GMS response
and the general detector response to the charged particles is
established, the GMS provides a way of monitoring the state
of the scintillator transmission length and phototube
gains. The GMS also proved to be useful in the times between
beam use periods, when configuration changes to the detector
and/or data acquisition system needed to be checked.

\section{Electronics and Data Acquisition}\label{sec:electronics}

The different requirements of the forward angle and backward angle
modes of the G0 experiment lead to rather different configurations
for the electronics used to process the signals from the detectors. 
As discussed in Section~\ref{sec:detectors},
the detectors and electronics
for octants 1,3,5, and 7 were built by a North American (NA) 
group, and those for octants 2,4,6, and 8 were built by a French
(FR) group. 
Because of differing backgrounds and expertise, significantly different solutions for the electronics are adopted by the two groups, particularly for the forward angle measurement.  In this section we first
summarize common aspects and the individual features of the forward angle electronics~\cite{Marchand}, followed
by a description of the common aspects of the backward angle configuration, again
supplemented by separate presentation of the two specific designs. 
Note that the use of two almost independent designs
for the electronics provides a valuable tool to search for false 
asymmetries or other artifacts that might be introduced by
problems in the electronics and that could thereby compromise   
the physics result of the experiment. The results of both the forward and
backward angle modes of the experiment 
are completely consistent between the 
the two sets of electronics, yielding no indication of such 
a problem (see, for example, Fig. 18 of Ref.~\cite{Marchand}). 

\subsection{Forward angle electronics}\label{forwardelectronics}

In the forward angle measurement, ToF from the target to the detectors is used to separate the elastic protons from inelastic protons, pions and other background.  Because of the high rates of a few MHz in each channel (FPD pair), time-encoding electronics (TEE) builds ToF spectra for each channel in hardware scalars; the spectra are stored by the data acquisition (DAQ) system using the standard readout controllers (ROCs) that are part of the Jefferson Lab CODA system~\cite{CODA} (see Section~\ref{DAQ}).

Signals from the detectors 
proceed first to a patch panel in Jefferson Lab Hall C through RG58 cables (total length of 36 m). From there they
 propagate to the electronics counting room through 107 m long
RG8 (for reduced attenuation) cables. At that point, the 
signals are split into two signals leading to the monitoring (Fastbus)
electronics and the TEE. The NA set-up uses
passive splitters, delivering 1/3 of the signal to the monitoring
electronics and 2/3 to the TEE. The FR
electronics uses active splitters with unity gain in both channels so
there is no loss of amplitude. 
%In both cases, the two outputs are completely decoupled.  
The monitoring electronics system is based on commercial ADC and TDC
units in Fastbus crates and is common to both the NA and FR
set-ups. It gives precise event-by-event information on the pulse
height and time response of the detectors, but is highly prescaled
to reduce the rate to a manageable level ($< 1$ kHz). 

The front-end of the TEE consists of
Constant Fraction Discriminators (CFDs), Mean Timers (MTs) and a
coincidence unit. Since off-line walk corrections are prohibited
by the lack of event-by-event data, CFDs are chosen in order to provide
good time resolution over a large dynamic range, the zero-crossing
being independent of the amplitude of the input signal. To reject
low-energy background, the CFD thresholds are set to 50~mV. The two
CFD output signals associated with the left and the right PMTs
of a given scintillator are mean-timed, so the
timing of an event delivered by the MT is independent of the hit
location on the scintillator paddle. When a coincidence between
the MT signals associated with the front and back scintillators of a given pair is
obtained, the event timing is encoded and the corresponding bin of
the ToF spectrum is incremented.

As mentioned earlier, the beam time structure of the G0 beam 
for the forward angle mode is chosen to be 31.1875 MHz 
(499 MHz divided by 16), and so there
is one beam pulse (``micropulse'') every 32~ns. 
The start of the ToF is generated by
a signal, $Y_0$, from the RF pulse of the electron beam
as it passes through a microwave cavity just upstream of the target~\cite{Musson}.  As detailed in Section~\ref{helicity_control},
the helicity of the beam is flipped at 30 Hz so
that one helicity state, referred to as one macropulse (MPS),
lasts 33~ms.
The readout of the
TEE data (ToF spectra) is performed at the end of each MPS,
during the helicity flip which takes about 500 $\mu$s. The selection of elastic events, as well as the calculation of all asymmetries, is performed during 
off-line analysis.

In normal data-taking the DAQ integrates each helicity state for 
two cycles of the AC line period, i.e., for 1/30~s. This cancels any noise
at the line frequency or its harmonics. To ensure that substantial
line noise is not present, occasional dedicated runs 
are performed in which each 1/30~s helicity state is integrated 
as four 1/120~s parts. These runs show negligible 60 Hz 
contributions. 

\begin{figure}[tbp]
\begin{center}
\includegraphics[width=4in]{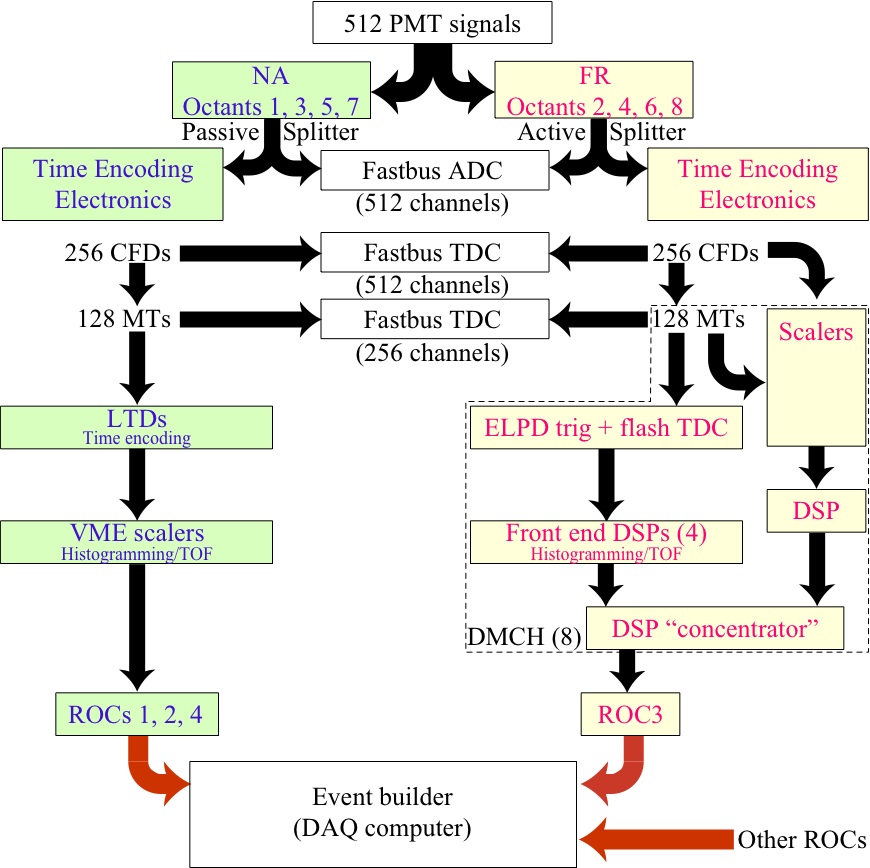}
\end{center}
\caption{Schematic of the forward angle detector electronics showing the NA and FR branches.  ToF is measured and recorded in different ways in the two cases (see text
%and Table~\ref{crates}
).}
\label{fig:Block diagram of the electronics set-up}
\end{figure}

As mentioned, two rather different solutions
are adopted by the NA and FR collaborations for the time-encoding 
electronics. The NA electronics
is modular and based on a combination of commercial
and custom-made elements. It is robust, but has limited binning
size (1~ns), due to the limitation of the maximum clock speed associated with the
technology employed.  The FR electronics, by contrast, 
is
highly integrated and has time resolution of 
250~ps, making some tasks, such as background correction,
easier. A block diagram showing the entire electronics chain, is
shown in Fig.~\ref{fig:Block diagram of the electronics set-up}.

The same two schemes are implemented in each design to reduce
helicity-dependent effects related to deadtime: the
Next-Pulse-Neutralization (NPN) and the ``buddy'' method. The
probability of one event being detected in a single 
micropulse is 3\% at 1 MHz. In order to allow the signals to
completely clear the mean-timer, the encoding is disabled for the
next micropulse (32 ns later). This increases the deadtime by
a few \%, but in a
controlled, deterministic way. The buddy method permits the study of
the deadtime for each detector by recording how often one detector
records a hit when its buddy (the same detector number located in 
the opposite octant of the detector array) is busy. 
This
quantity is monitored in particular for indications of helicity-correlated
structure in beam intensity, which would be seen as a
helicity-correlated variation in these buddy rates. This permits
us to monitor for helicity-correlated deadtime losses which might
introduce a false asymmetry. For the FR electronics it is also
possible to use the buddy method for each bin of a ToF spectrum,
{\it{e.g.}} for the elastic peak (``differential buddy").

\subsubsection{NA Electronics - forward angle}

The NA electronics system uses
separate modules to perform the tasks of discrimination,
meantiming, time-encoding, and ToF spectrum accumulation.
Additional modules are used to generate the clocking signals used
by the time-encoding boards.

Commercial LeCroy 3420
CFDs~\cite{LeCroy} are used to
minimize time-walk in the PMT anode pulses.  Custom
mean-timers, utilizing the same application-specific integrated circuit (ASIC) as in the FR electronics, are then used to average the pulse times of the PMTs at
the opposite ends of each scintillator.
The ASICs introduce the two signals into counter-propagating
shift-registers and generate a mean-time signal when a coincidence
occurs.  Copies of the CFD
outputs and of the mean-timer outputs are also sent to the monitoring
electronics. 

\begin{figure}[tbp]
\begin{center}
\includegraphics[width=4in]{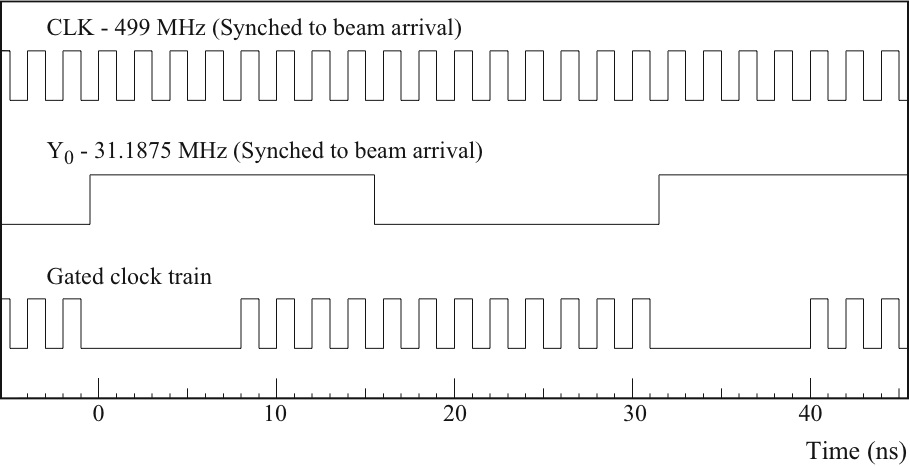}
\end{center}
\caption{Timing diagram showing gated clock trains generated
based on the free clock (CLK) and beam arrival ($Y_0$) signals for the NA electronics, as 
described in the text.}
\label{fig:clock_train}
\end{figure}

Mean-timed signals from the focal plane detectors are sent to the
time-encoding boards, the ``Latching Time Digitizers'' (LTDs).
First, a coincidence is required between the front and 
corresponding back detectors, with the timing being determined by
the front detector.  The ToF is determined using a clocked shift register,
a very simple method for accumulating time spectra for
data rates of several MHz.  With an overall cycle time of 32~ns, an
externally generated train of 12 clock pulses, synchronized to the $Y_0$  signal (see 
Fig.~\ref{fig:clock_train}) is used to clock a shift register whose
input is latched on by the front-back coincidence.
The depth of penetration of the input signal into the shift
register during the shifting sequence then depends upon the time
of the coincidence within the 32 ns cycle.  The depth of
penetration of the signal thus encodes the time of the
coincidence.  A time spectrum can then easily be recorded by
presenting the parallel output bits from the shift registers 
to individual scaler channels.  Scaler channels corresponding to times
subsequent to a coincidence are incremented while those
corresponding to earlier times in the 32 ns cycle are not.  The
shift register input is latched through the remainder of the clock
train to simplify extraction of the time spectrum from the scaler
information by taking differences of successive scaler channels.
After a period of accumulation, each of these differences
represents the number of coincidences which came within a time bin
early enough to increment one scaler channel but not early enough
to increment the next channel.

The CLK and $Y_0$ signals are sent to a custom board, the
clock-gating board, which gates off a group of 4 pulses of CLK,
synchronized to $Y_0$.  The resulting 12-pulse clock trains are
duplicated by custom signal-duplication boards and distributed to
the LTDs to act as the time base for ToF measurements.  Restricting
the clock train to 12 pulses allows us to record only the times of
interest (starting at the time when the fastest particles reach the
detectors).

The LTD boards incorporate a few refinements of the basic
mechanism described above.  In fact, the latched input signal is
presented to the inputs of two shift registers, one of which is
clocked by the leading edges of the clock train and one of which
is clocked by the trailing edges.  This effectively halves the
time bin size since interleaved differences can be taken between a
scaler channel on one shift register and one on the other.  The
clock pulse frequency is 499 MHz, so this interleaving reduces the
time-bin width from roughly 2~ns down to roughly 1~ns.  The input
latch is not simply reset at the end of each clock train. Rather,
if it is set in a particular clock train, it is cleared at the end
of that clock train and also disabled for the duration of the
subsequent clock train.  As noted above, this enforced, extended
deadtime is intended to make deadtime corrections more accurate by
reducing dependence on the less well-defined deadtime properties of
PMTs, CFDs and mean-timers.  Each LTD encodes times from two front
scintillators (buddy pairs).

\subsubsection{FR Electronics - forward angle}

The FR electronics
consists of eight custom mother boards called DMCH-16X
(Discrimination, Mean-Timing, time enCoder, Histogramming, 16
Mean-Timer channels within the VXI standard), each handling the eight detectors (32 PMT channels) of half an octant.  Because of integration, the electronics for four octants fits in a standard, C-sized VXI crate.  An interface module
provides common signals (MPS, $Y_0$) to the DMCH-16X boards through
the VXI back-plane. One DMCH-16X board
consists of

\begin{itemize}
\item 32 CFDs\footnote{The CFDs use integration-differentiation type shaping,
relying only on an RLC cicuit, not delay lines.  The CFD thresholds are controlled by software.} and 16 MTs grouped into 16 CFD-MT daughter boards, each of them holding 2 CFDs and one MT, 

\item 4 EPLD-Trig (Electrically Programmable
Logic Device for Trigger) modules dedicated to logic, in particular the coincidence between front and back MT signals, 

\item 2 custom numerical time encoders locked to the
$Y_0$ signal, 

\item 4 asynchronous First-In-First-Out (FIFO) buffers (2048 words) to hold the events on the way from the time encoders to the front end Digital Signal Processors
(DSPs), 

\item 4 front-end DSPs (ADSP-21062 SHARC~\cite{AnalogDevices}) for histogramming,

\item a Field Programmable Gate Array (FPGA) chip and a DSP grouped onto a daughter board (``SDMCH'') to provide individual CFD and MT scalers independent of the ToF data, 

\item 1 DSP concentrator (ADSP-21062 SHARC) which collects data from the 5 DSPs, and

\item an internal signal generator (``GDMCH'') for testing the CFDs and MTs.
\end{itemize}

\begin{figure}[tbp]
\begin{center}
\includegraphics[width=5.5in]{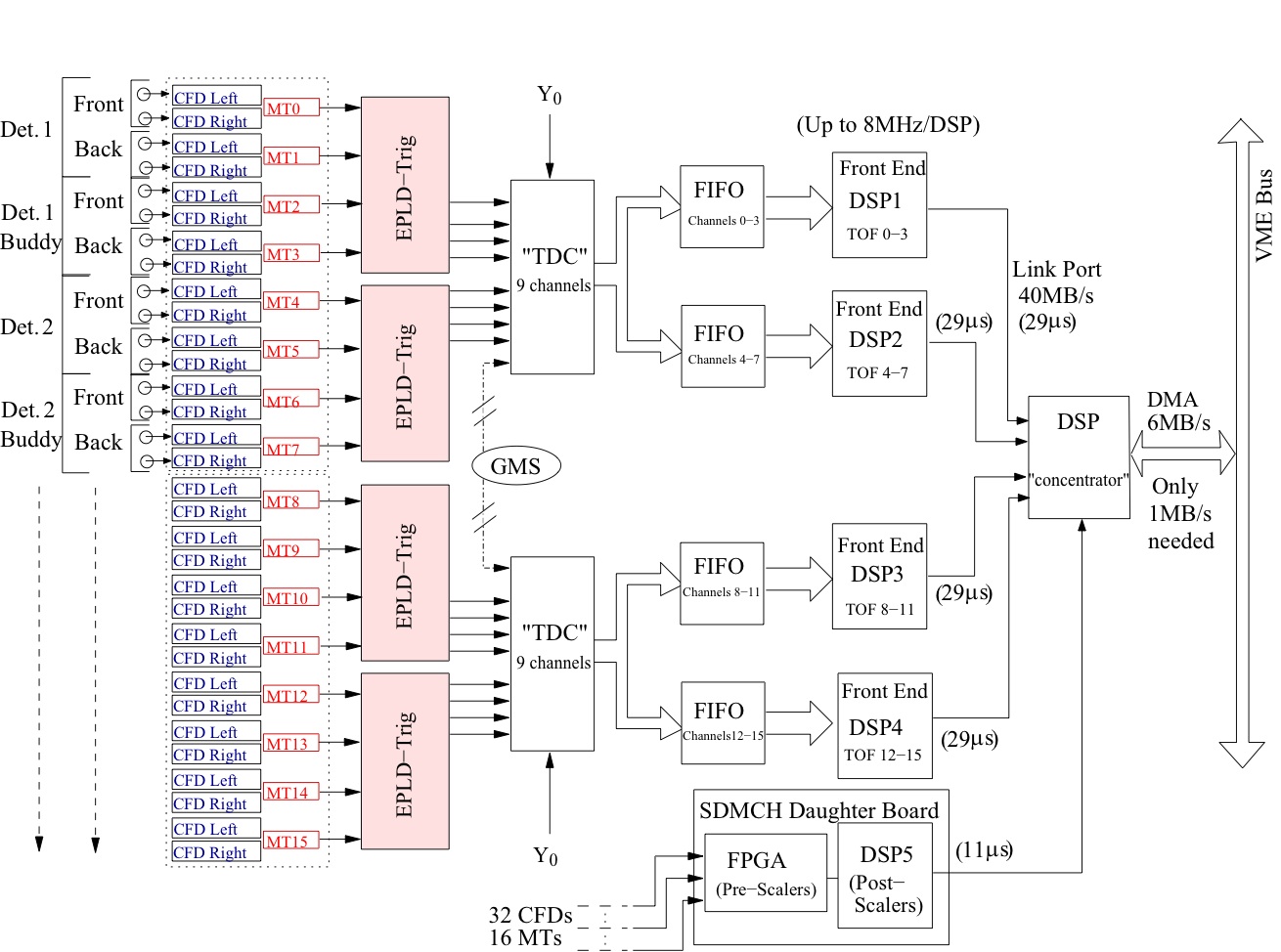}
\end{center}
\caption{Architecture of the FR DMCH boards for the forward
angle measurement. Each board 
processes 32 PMT signals corresponding to 8 pairs of
scintillators.}
\label{fig:dmch_architecture}
\end{figure}

A schematic of the architecture of the DMCH-16X mother board is
presented in Fig.~\ref{fig:dmch_architecture}.

The processing sequence begins as each CFD, after an input signal, is disabled until the end of the mean-timing sequence. Based on the longest
scintillator paddle, the MT compensation range is set to
17 ns; the overall MT deadtime is about 37 ns. 
The front MT signals are used to set the overall timing. 
To achieve better time matching between the front and the
back MT signals, a software-controlled, internal delay
can be adjusted. This delay can be set 
from 0 to 44 ns in 0.175 ns steps. 
The coincidence
window, which is generated by the EPLD-Trig once a back MT signal
arrives, is set to 7~ns. Apart from the coincidence logic,
other modes have been implemented in the EPLD-Trig chip that
can be selected by software for testing purposes. These test modes
allow one to build ToF spectra of only front signals, only back
signals, or of both. The NPN and the Buddy schemes can
also be disabled for testing purposes. 

The numerical time encoder is an ASIC with
250~ps time resolution and a start time (the $Y_0$ reference signal) input.
It utilizes a slow and a fast counter to achieve the time resolution.
Briefly, after appropriate frequency division, a phase-locked loop is
used to lock the frequency ($\sim250$~MHz) of a Voltage Control
Oscillator (VCO) onto the $Y_0$ signal. A slow counter
running at the VCO frequency divides the 32~ns
period of the $Y_0$ signal into $\sim 4$~ns periods. A fast counter
consisting of 8 delay-locked loop circuits then divides the
4~ns period into 16 bins of 250~ps. 
Once the time is flagged, the
information is stored in one of the 9 independent coding
registers, and then transferred to the corresponding storage
register using a synchronous FIFO buffer. The intrinsic deadtime of the
time encoder is 24 ns and therefore does not introduce additional deadtime beyond that of the MT. Because of the design and the technology of the ASIC, the time encoder has a significant intrinsic differential non-linearity, which is corrected off-line.

In order to sustain a mean rate of a few MHz per channel (scintillator pair), the code of the 4 front-end DSPs is optimized so
that a ToF bin is incremented using only 4 instructions (5 cycles
of 25~ns). Consequently, the maximum rate the DSP can handle is
8~MHz. As two such DSPs are associated with one time encoder in the normal
coincidence mode of operation, the maximum rate per detector is
4~MHz.

Independent of the time encoding data, an additional daughter board
(SDMCH) containing an FPGA chip and a DSP 
provides individual scalers for the CFD and MT signals. Complementary to the
Fastbus monitoring data, this scaler information helps in
quantifying the deadtime associated with incomplete events, such as
single CFD hits. The principle of these scalers is based on pre-scalers
implemented on the FPGA chip and post-scalers collected by the SDMCH
DSP.

At the end of each MPS, the 4 front-end DSPs and the SDMCH DSP rapidly
transfer their data to the DSP concentrator through link ports
(40~MB/s) working in parallel. This transfer takes 29~$\mu$s during
the 500~$\mu$s dedicated to the beam helicity flip. The data transfer
(5~kB) from the DSP concentrator to the ROC occurs during the next
MPS using Direct Memory Access (DMA). At a rate of 6~MB/s, the
transfer lasts 16~ms. Via the FR ROC, the DMCH-16X data are
gathered with the data from other ROCs in the CODA Event Builder.
The total flow of the FR data is 1.4~MB/s which represents 2/3 of
the total G0 data transfer.

\subsection{Backward angle electronics}
For the backward angle measurements, a coincidence is formed between the signals from the FPDs and those 
from the CEDs mounted at the exit of the 
magnet, effectively determining
the scattered electron momentum and angle, and thereby
separating the elastic and inelastic electrons.  Because ToF is not measured, the backward angle trigger is formed from a coincidence of the OR of all CEDS in an octant with an OR of the FPDs in the same octant.
In addition to the signals from the
CEDs and FPDs, \v Cerenkov detectors are employed, which 
generate a signal for the scattered electrons, but do not
generate a signal from background $\pi ^-$'s. The signal from the
\v Cerenkov detector is therefore used to enable the
coincidence logic.  The required coincidences 
are performed in custom electronics using
programmable logic devices. 
In contrast to the forward measurement, the coincidence rate is only $\sim 100$ kHz for the elastic peak.  

\begin{figure}[tbp]
\begin{center}
\includegraphics[width=5.5in]{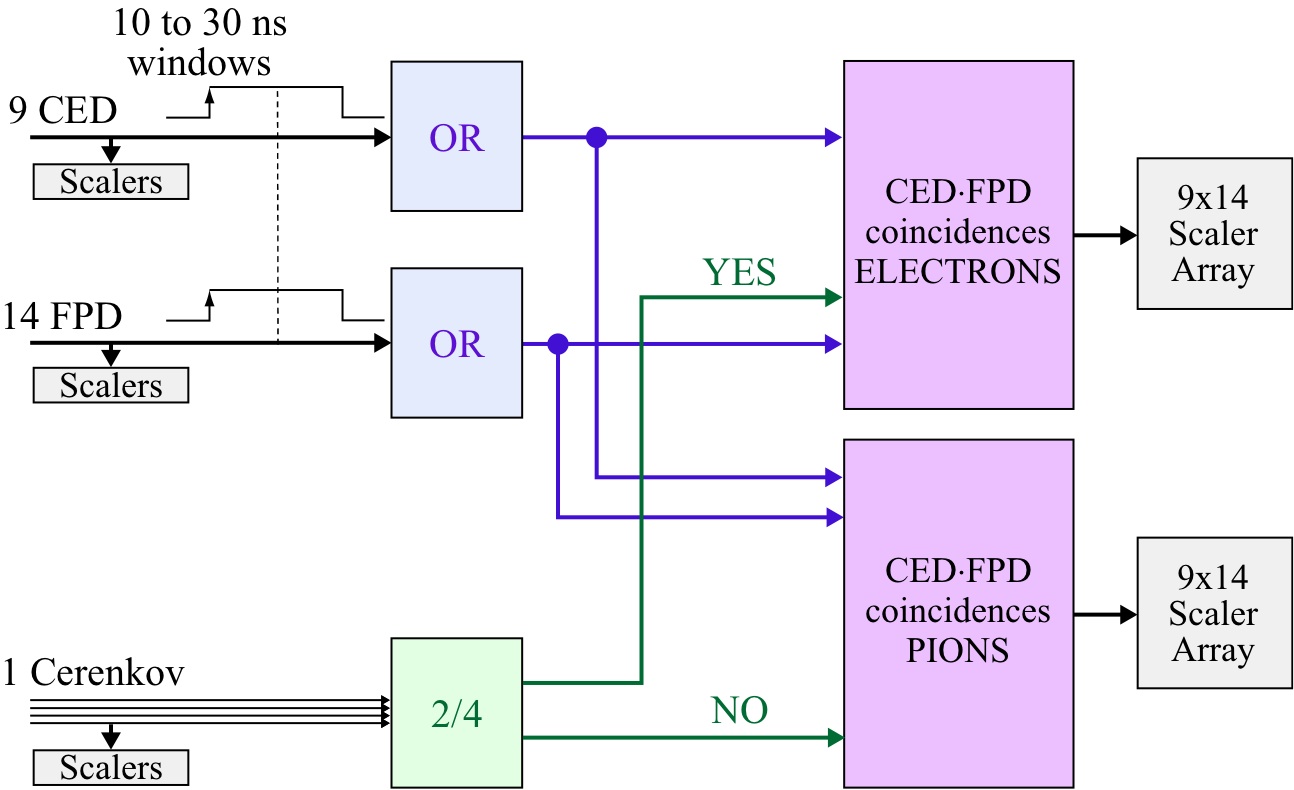}
\end{center}
\caption{Schematic and timing diagram for the backward angle
electronics.  The NA and FR layouts are similar, but with different components for the coincidence and scaler modules (see text).}
\label{fig:BA_electronics}
\end{figure}

A summary of the basic logic in the backward angle electronics can be
found in Fig.~\ref{fig:BA_electronics}. 
The CED, FPD, and \v Cerenkov signals
are combined to form the CED-FPD
coincidences.  Both coincidences with (electrons) and without (pions) the \v Cerenkov signal are recorded.  Some data were also recorded (in the ``pion'' scalers) with the \v Cerenkov signal delayed by $\sim 100$ ns to provide a direct measure of the random coincidences.  The logic is also configured to reject events having
more than one CED signal, or more than one FPD signal
for a given beam burst. These multiple hit signals,
as well as signals for both CED and FPD singles events
are
recorded to assist in deadtime loss and pile-up corrections.  Again, implementations which differ in detail are adopted for the NA and FR octants, and they are described 
separately in the following. 

\subsubsection{NA electronics - backward angle}

In the backward angle NA electronics, these logic functions are performed
in a series of custom boards which replace the LTD 
boards used for the forward angle
measurements. For each of the four NA octants, the 
outputs of
the CED and FPD mean timer modules are split, sending one copy to
the scalers, and one copy to the main 
logic board. 
The main logic board
performs the timing coincidence between the CEDs and FPDs,
determines if a multiple hit event occurred, and sorts the
coincidences according to which CED-FPD pair is hit.
An 
Altera programmable logic device (PLD)~\cite{Altera} performs the logic
functions. 
The CED-FPD
coincidence information, for events without multiple hits, is then encoded 
into one 8-bit word, and
sent on the crate back plane to a set of three custom boards which
decode the information to a form suitable for scaler input.

In addition to the 8-bit word, one
bit indicating the existence of a multiple hit event, and all CED
and FPD singles events which had the correct timing, are sent on
the back plane to another custom board which again uses a
PLD to perform a coincidence between the CED
and FPD correct-time singles events with the multiple hit bit.

The 8-bit
word containing the CED-FPD coincidence information is taken from
the back plane by the 
custom decoder boards and latched in
another PLD to avoid timing jitter among the 8 separate
signals comprising the word. 
The \v
Cerenkov signal is also used as an input to the decoder PLDs to
release, as appropriate for electrons and for pions, the information through the latch. The 8-bit word is
then decoded into individual CED-FPD coincidence signals, which
are 
sent to the scalers for recording.
To accommodate the four NA octants, a custom crate with 4
separate 5-slot back planes is employed. \\

\subsubsection{FR electronics - backward angle}

The FR electronics design for the backward angle measurement makes use of a part of the
custom module developed for the forward angle measurement. As in the NA electronics, it is based on
the coincidences between CED and the FPD detectors, enabled by the \v Cerenkov signal (corresponding to the
detection of electrons); see Fig.~\ref{fig:BA_electronics}.
Each of the 9$\times$14 CED-FPD
coincidences, for both the electron and pion cases, are then histogrammed during the MPS and
transferred to the acquisition system during the helicity
reversal time. The CFDs, mean-timers and scalars for the forward angle
measurement are re-used, in conjunction with an additional VXI module
based on an Altera PLD~\cite{Altera}.  In a manner similar to that in the NA electronics, the logic board also provides capabilities for recording singles and multihit events used to characterize the deadtime.

\subsection{Monitoring electronics}\label{monitoring}                                             %1 page           41.5
As described above, custom electronics are required to accumulate
data at the very high rates needed to give sufficient statistics
for precise measurement of the small asymmetries of interest in
this experiment.  The trade-off for making measurements at such
high rates is that very little information can be recorded, as
event-by-event recording is prohibitive (data transfer and recording rates).  A separate set of
conventional LeCroy Fastbus ADCs (1885F) and TDCs (1875A)~\cite{LeCroy}
is used to capture much more detailed event-by-event information
for a tiny fraction of the events.  For the forward angle, the trigger used to start
the TDCs and gate the ADCs is simply a pre-scaled version of the
$Y_0$ signal; in the backward angle case a conventional start/gate is
provided based on the pre-scaled CED-FPD coincidence trigger described above.  The pre-scale value is chosen to reduce the trigger rate
to, typically, a few hundred Hz.  During normal operation, the collection, digitization, and
readout of a monitoring event takes about 1 ms.

Splitters are used to send a copy of each PMT signal to an ADC
channel as well as sending it to the discriminator input.
A copy of each CFD output is sent to a
TDC channel, as is a copy of each mean-timer output. In addition, for the backward angle data, a simple wave form digitizer (``analog ring sampling''
module) is used to monitor the analog signals from the individual Cerenkov PMTs.  It samples the signals at 1 GHz over 128 ns for each monitoring event.  

These monitor
electronics make it possible to check the basic operation of the
PMTs, discriminators and mean-timers. Furthermore, the correlated
event-by-event information allows investigation of a wealth of
effects which could not be seen with only the data from the main
high-rate electronics.  These include monitoring of the rate, time-distribution and pulse height of single PMT hits (to check PMT gains, discriminator thresholds, etc.), single scintillator hits,
etc.

\subsection{Data acquisition electronics}\label{DAQ}                                        
The G0 data acquisition system is built on CODA~\cite{CODA}, developed at Jefferson Lab. The
electronics subsystems each occupy one or more crates, with
triggering and event control performed by the Trigger Supervisor
module and the Trigger Interface modules~\cite{Jas99}.

% A summary of the control layout for the DAQ system is given in %Table~\ref{crates}.

%\begin{table}[tbp]
%\begin{center}
%    \begin{tabular}{p{1.2in}ccp{2in}}
%    Source & Controller & Crate & Modules\\
%    \hline
%    NA electronics      & ROC1  & VME   & Scale-32\\
%                & ROC2  & VME   & Scale-32\\
%                & ROC4  & VME   & Scale-32\\
%    FR electronics  & ROC3  & VXI   & DMCH-16X\\
%    Monitor electronics & ROC5  & Fastbus
%                    & LeCroy 1875 \& 1885 modules\\
%    Beam and control electronics
%    & TS0   & VME
%                    & TS, I/O register, SIS3801,
%                      Multihit TDC, CAMAC interface,
%                  STR7200\\
%    \end{tabular}
%\end{center} 
%\caption{Layout of the DAQ electronics.}
%\label{crates}
%\end{table}

The standard event, monitoring, and GMS trigger sources are input to the Trigger Supervisor.  
It registers the arrival of the trigger signal,
and begins processing the event.  The trigger type information is
passed to the other crates through a connection between the
Trigger Supervisor and the Trigger Interface module in each crate.
The DAQ software for each crate reads out individual modules
depending upon the trigger type.  For example, for monitor events, the Fastbus crate is read out through ROC 5, but ROCS 1-4 and the beam and control electronics
are not read out.

As is mentioned in Section~\ref{monitoring}, the trigger
source for the monitoring events is a prescaled copy of either the $Y_0$ signal (forward angle) or coincidence signal (backward angle).
It is prescaled in hardware to provide a trigger input rate of $<1$ kHz. There is an additional prescale in the Trigger
Supervisor software; during typical operating conditions, 
the overall accepted trigger rate is a few hundred Hz.

We normally read out information at 30 Hz, following each MPS. In the 120 Hz over-sampling mode, there are two trigger
types used: A, which occurs at 30 Hz, and B, which corresponds to
the three additional 120 Hz phases between each MPS trigger. The FR electronics is read for type A; the NA for both type A and B. 

\subsection{Data analysis}                                              

A brief description of the analysis system is given here to complete
the overall description of the G0 apparatus.  The main G0 replay code has, as input, a CODA file
produced by the DAQ system.
Its primary outputs are histograms and ntuples, filled on an
``event-by-event'' (MPS or QRT) basis, and a MySQL~\cite{MYSQL} database where
only values averaged over a run (typically one hour in  
duration) are saved. The database is used as the primary 
source for further
analysis, whereas the histograms and ntuples are mainly used for
analysis of calibration data. For each of the
forward and backward angle runs, roughly 
10 TByte of data are recorded in CODA files, and
these raw data are subsequently reduced to a database of $\sim 10$ GByte.

\begin{figure}[tbp]
\begin{center}
\includegraphics[width=5.5in]{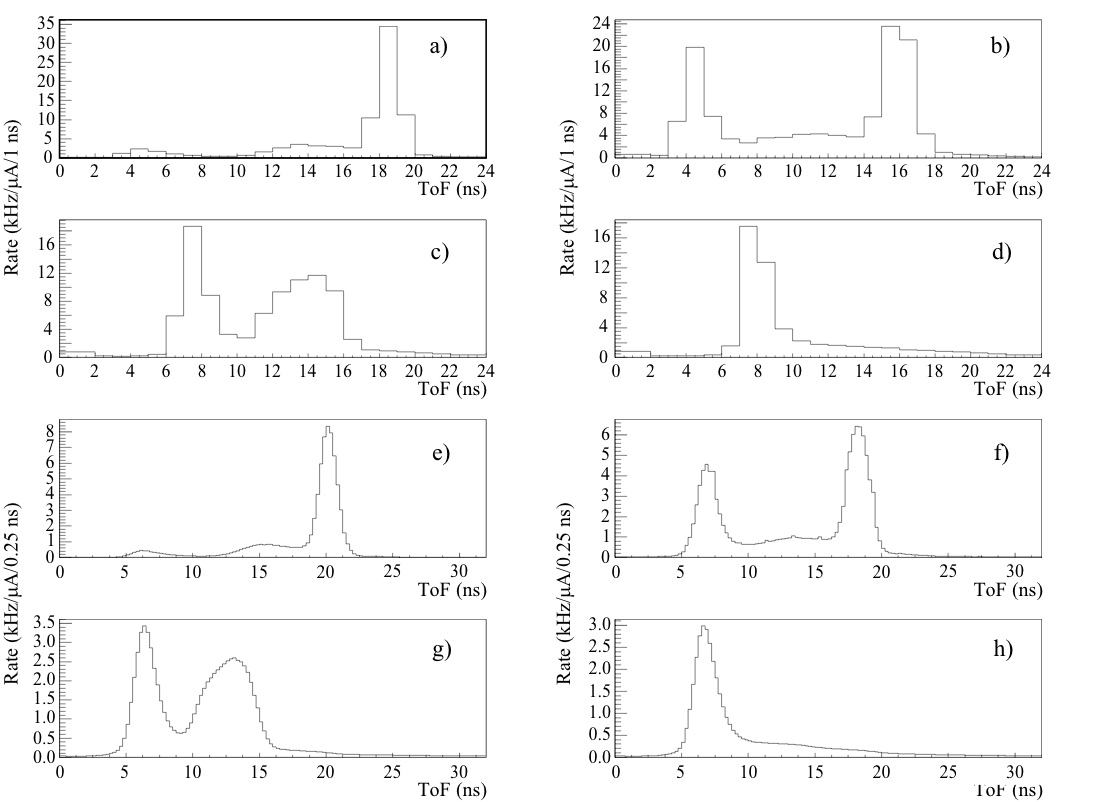}
\end{center}
\caption {Sample forward angle
ToF spectra: a) NA FPD 1, b) NA FPD 11, c) NA FPD 15, d) NA FPD 16, e) FR FPD 1, f) FR FPD 11, g) FR FPD 15, and h) FR FPD 16 ($t=0$ is slightly different for the NA and FR detectors). For detector numbers 1 to 14, the
elastically scattered protons appear as a sharp peak at late
ToF, while the pions appear at earlier ToF,
with their measured rate increasing with increasing ring number.
For detector 15, the elastically scattered protons appear as a broad
distribution because the Q$^2$ range of the acceptance of this
ring is $\sim0.4$ (GeV/c)$^2$ compared to less than 0.1 (GeV/c)$^2$ for
the other detectors. Finally, there is no acceptance for elastic
events in detector 16, where only pions and inelastically scattered
protons are measured.  Note that for $24<t<32$ ns, for which we recorded information only with the FR electronics, the count rate is negligible.} 
\label{fig:tof}
\end{figure}

\begin{figure}[tbp]
\begin{center}
\includegraphics[width=5.5in]{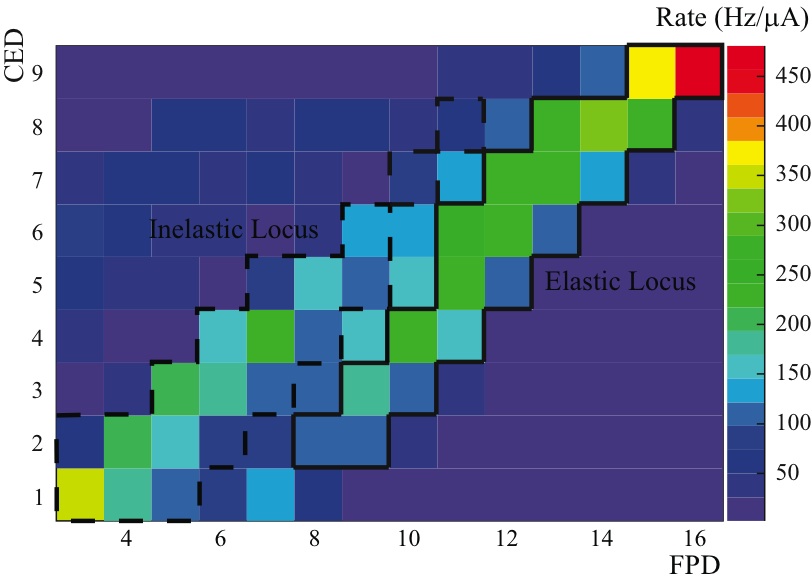}
\end{center}
\caption {Backward angle measurement of scattered electrons from the LH$_2$ target
at 684 MeV incident energy, showing coincidence rates for the various combinations of CEDs and FPDs.  The elastically scattered electrons appear in a band toward the upper right, the inelastically scattered electrons in a band toward lower left as shown.} 
\label{fig:matrix}
\end{figure}

The main analysis treats the MPS (30~Hz) events which are of two types.
For the forward angle measurement, the first type consists of
ToF spectra associated with the FPDs (see
Fig.~\ref{fig:tof}); for the backward angle it comprises the matrix of CED-FPD coincidences (see Fig.~\ref{fig:matrix}). 
For both cases, the second type consists of
beam diagnostic data, integrated over the 33~ms helicity window.
These data include measurements of beam current, energy, position and 
angle, and the data from the luminosity monitors.

The analysis of the 30 Hz data is based on the following
procedure. For each MPS and for each ToF (forward) or coincidence (backward) channel, the yield ($N$)
is normalized by the beam charge
accumulated during the MPS. The event is tagged as
bad if the average beam current on target during its MPS is less
than (typically) 5$\mu$A, or if the electronics associated with this detector has sent an error bit. The helicity state (which is encoded in
the data stream with a delay of 8 MPS) of the beam is then decoded
and associated with each MPS event.  
As discussed in Section~\ref{helicity_control}, the beam helicity 
sequence is produced in a quartet structure of macropulses.
The first MPS of
each quartet is tagged in the data stream, and if the analysis has
identified all four of the MPS in the quartet to be good, an
asymmetry is computed for each channel

\begin{equation}\label{eq:asymmetry}
A({\rm quartet, time bin}) =
\frac{\left(\frac{N_+^1}{Q_+^1}+\frac{N_+^2}{Q_+^2}\right)-
\left(\frac{N_-^1}{Q_-^1}-\frac{N_-^2}{Q_-^2}\right)}
{\left(\frac{N_+^1}{Q_+^1}+\frac{N_+^2}{Q_+^2}\right)+
\left(\frac{N_-^1}{Q_-^1}+\frac{N_-^2}{Q_-^2}\right)} \;\;\;\; ,
\end{equation}

where $N_{s}^i$ is the number of counts recorded for the $i^{th}$
MPS of this quartet with beam helicity of sign $s$, and $Q_s^i$ is
the beam charge incident on target during this MPS. If one or
more of the MPS of the quartet is tagged as bad, the quartet is
not included in the main run-averaged information.  The asymmetries for
a given target/angle combination are
multiplied by a blinding factor (0.75 to 1.25) at this stage, and ``unblinded''
at the completion of the analysis.

For each channel, the asymmetry, as
computed above, is averaged over the run with an uncertainty
given by the root mean square deviation of the asymmetries over
the run divided by the square root of the number of good
macropulses for that channel. These values are saved in
the database.  Beam charge asymmetries and beam position
differences are computed following the same algorithm of quartet
identification and averaged for a run.

Corrections for three main effects, counting rate, helicity-correlated beam changes and backgrounds, are also computed in the analysis software.  The rate corrections, both for deadtime and for random triggers are electronics-dependent and therefore different for the FR
and the NA data~\cite{Liu,Versteegen}. Measured electronics busy fractions are combined with concurrent measurements of singles rates and asymmetries to make corrections to the raw asymmetry.  These corrections typically amount to a few percent of the measured asymmetries.

Forming asymmetries using rate-corrected yields
constitutes a ``first-pass'' analysis. During this pass, the
sensitivity of each detector channel to variations in the beam
 position, angle, energy and current (the
``yield slopes'' $\partial Y / \partial P_{i}$, 
see Section~\ref{sec:beam_mon}) are also computed.  In a second pass through the analysis,
the detector yields $N^i_s$  are corrected using these yield slopes and
the corresponding, measured, helicity-correlated beam properties,
prior to re-computing
the asymmetry (Eqn.~\ref{eq:asymmetry}). 
It should be
noted that the correction for yield variation as a function of the helicity-correlated beam current changes is small, both because the beam current
changes themselves are small and because the first-order effects have already been removed by the (helicity-averaged) rate corrections.

\begin{figure}[tbp]
\begin{center}
\includegraphics[width=5.5in]{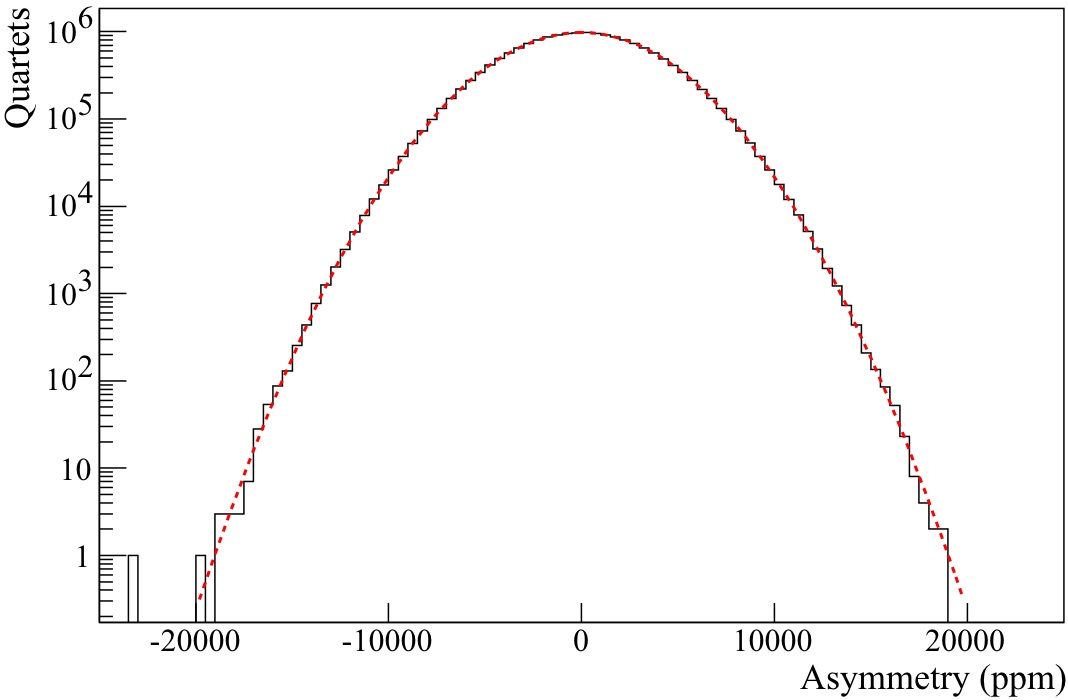}
\end{center}
\caption{The distribution of the measured 
forward angle, octant 7, FPD 8 proton asymmetries ($1.7\times10^7$ quartets) together with a Gaussian fit ($\chi^2/\nu = 84.2/74$).}
\label{fig:gaus}
\end{figure}

\begin{figure}[tbp]
\begin{center}
\includegraphics[width=5.5in]{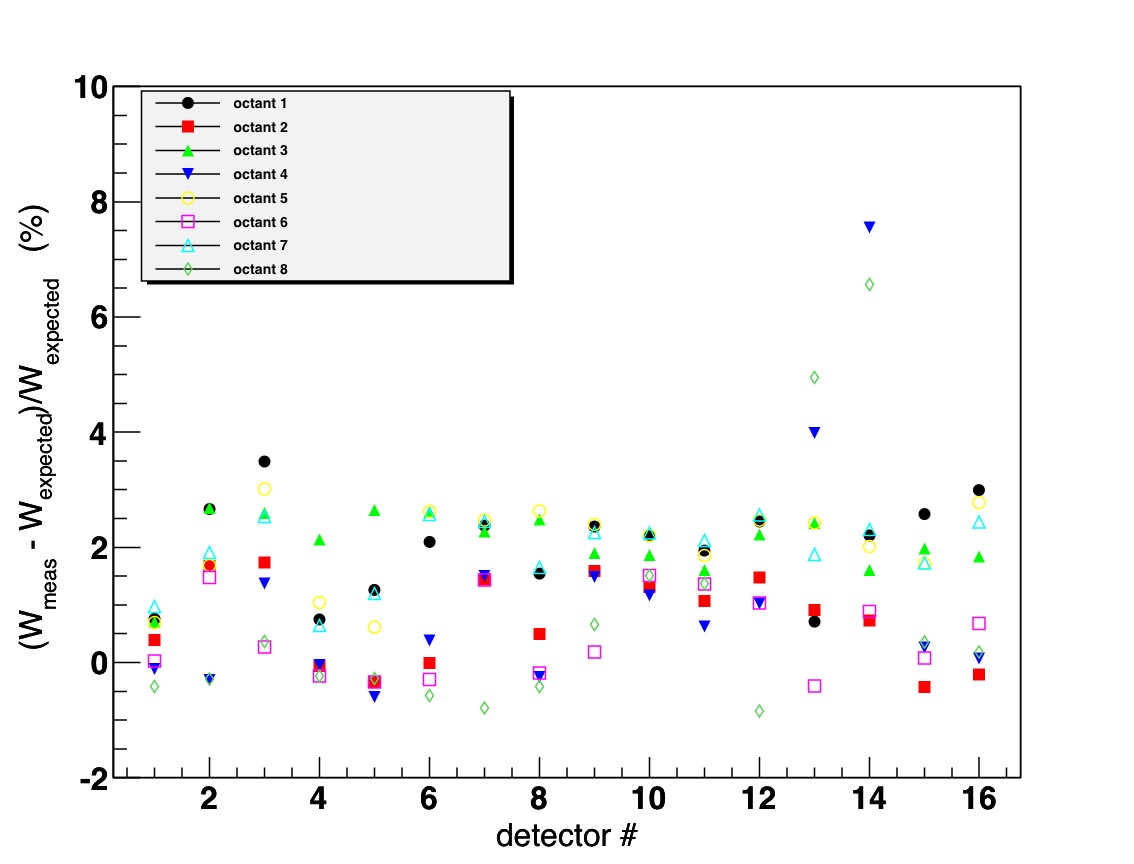}
\end{center}
%\caption{The ratio of the observed and statistically 
%expected forward angle proton asymmetry widths for all FPDs in all octants.}
\caption{The percentage difference between the measured and statistically 
expected forward angle proton asymmetry widths for all FPDs in all octants.}
\label{fig:width}
\end{figure}

An example of the resulting distribution of the QRT asymmetries for one FPD pair (for the entire forward angle measurement) is shown in Fig.~\ref{fig:gaus}, showing the expected Gaussian distribution with no tails. 
 The expected width of this distribution can be computed from the measured count rate, after correction for dead-time effects~\cite{Guillon};
Fig.~\ref{fig:width} shows the ratio of the
standard deviation of the measured distribution (forward angle) to that expected from the rates. 
We measure a distribution only $\sim 2$\% broader than that expected from counting statistics. As discussed in 
Section~\ref{targetperformance}, a broadening of the width of order 1--2\% is expected due to target density
fluctuations. The few detectors with somewhat larger widths are likely either due to some additional electronic
noise or an underestimate of the electronic deadtime. Neither effect is expected to be helicity-correlated, so would not 
introduce a systematic error, only a modest decrease in the statistical precision.

Finally, starting from the run averaged information contained in the database, corrections for backgrounds are made using both the yield and asymmetry data, typically from neighboring channels and/or different triggers acquired concurrently with the main data.

\section{Summary and Operation History}
\label{sec:summary}

The major components of the G0 experiment have been described. In addition to the experiment-specific hardware provided
by the collaboration, such as the superconducting toroidal
spectrometer, cryogenic target, and detector system, we have provided
a description of the polarized electron beam produced by the CEBAF
accelerator.

The G0 experiment was first installed in Hall C of Jefferson
Lab during the period July 2002 to October 2002. The first
commissioning run took place over the next three
months. In addition to commissioning the various hardware
components associated with the experiment, significant time was
spent by the accelerator group developing the specialized beam
required by the G0 experiment; i.e., the modified time structure, the
high beam current with the corresponding large beam bunch charge, and 
the required small helicity-dependent beam properties.
%At
%the end of the commissioning run, all aspects of the experiment
%met the required specifications, or the necessary
%modifications required to meet the specifications were understood.

The G0 experiment was reinstalled, aided by the fact that the toroidal
magnet and detector system were mounted on rails for easy
installation and removal, during the Fall of 2003. Another
commissioning/engineering run followed installation. The actual
production runs for the forward angle measurement began about
mid-March 2004 and continued until about May 2004, at which time
the statistical goals of the experiment had been met. The results
of the forward angle run have been published as a letter~\cite{PRL05}. A longer paper providing details is in preparation.

Having completed the forward angle measurement, the new hardware
for the back angle measurements was installed in
Hall C, and commissioning and initial data-taking took place in 
March to May of 2006.  A long period of production running followed -- including measurements with both hydrogen and deuterium targets, each at beam energies of 359 and 684 MeV -- ending in April 2007.  The results of the main backward angle electron asymmetry measurements have been published~\cite{Androic}; a longer paper providing details about this phase of the experiment is also in preparation.

\section{Acknowledgments}
\label{sec:acknowledgements}

This work is supported in part by grants from several agencies.  The U.S. DOE provided support for the construction of the North American detectors, the North American electronics, for the infrastructure and operation of Jefferson Lab, as well as for the activities of some collaborating groups.  The U.S. NSF provided funding for the superconducting magnet and the cryogenic target, as well as for the activities of some collaborating groups.  The French CNRS provided support for the French detectors and electronics, as well as for the activities of the French collaborators.  The Canadian NSERC provided support for the North American detectors and electronics as well as for the activities of the Canadian groups.  We thank all the funding agencies for their support of the experiment.

We would like to thank the administration and staff of Jefferson Laboratory, especially members of the Accelerator and Hall C groups who made strong contributions to the success of the experiment.  We would also like to acknowledge the contributions of, and to thank the technical staff at Caltech (target), Carnegie Mellon (detectors, electronics), Grenoble (detectors, electronics), Illinois (magnet), Orsay (detectors, electronics), and TRIUMF (detectors, magnet).

\end{document}